\documentclass[11pt]{article}
\usepackage{amsmath,amsfonts,sint,epsfig,cite,graphics}
\usepackage{amssymb,bbm}
\usepackage{mathrsfs}  
\usepackage{amsbsy,color}
\usepackage{latexsym}
\usepackage{graphicx}
\usepackage{psfrag}
\usepackage{url}
\newcommand{\be}{\begin{equation}}
\newcommand{\ee}{\end{equation}}
\newcommand{\bea}{\begin{eqnarray}}
\newcommand{\eea}{\end{eqnarray}}
\newcommand{\ba}{\begin{array}}
\newcommand{\ea}{\end{array}}

\newcommand{\eq}[1]{eq.~(\ref{#1})}
\newcommand{\Eq}[1]{Eq.~(\ref{#1})}

\newcommand{\Eqs}[1]{Eqs.~(\ref{#1})}
\newcommand{\fig}[1]{Fig.~\ref{#1}}
\newcommand{\Fig}[1]{Figure~\ref{#1}}
\newcommand{\sect}[1]{Sect.~\ref{#1}}

\newcommand{\app}[1]{App.~\ref{#1}}

\newcommand{\tab}[1]{Table~\ref{#1}}
\newcommand{\Tab}[1]{Table~\ref{#1}}
\newcommand{\Ref}[1]{Ref.~\cite{#1}}

\newcommand{\p}{\phantom{0}}

\newcommand{\off}[1]{ }

\newcommand{\tauint}{\tau_\mathrm{int}}
\newcommand{\tauintl}{\tau_\mathrm{int}^\mathrm{l}}
\newcommand{\tauintu}{\tau_\mathrm{int}^\mathrm{u}}
\newcommand{\tauexp}{\tau_\mathrm{exp}}
\newcommand{\Obar}{\overline{O}}
\newcommand{\Gammabar}{\overline{\Gamma}}
\newcommand{\Sigmabar}{\overline{\Sigma}}
\newcommand{\nf}{N_\mathrm{f}}
\newcommand{\rmO}{\mathrm{O}}
\newcommand{\Oa}{\mathrm{O}(a)}
\newcommand{\Wl}{W_\mathrm{l}}
\newcommand{\Wu}{W_\mathrm{u}}
\newcommand{\Force}{{\mathcal F}}

\newcommand{\simas}[1]{\raisebox{-.1ex}{
            $\stackrel{\small{#1}}{\sim}$}}

\begin{document}

\begin{titlepage}

\begin{flushright}
\small{
DESY 10-151 \\
SFB/CPP-10-81\\
HU-EP-10/55\\}
\end{flushright}

\begin{center}
{\Large\bf
Critical slowing down and error analysis in lattice QCD simulations
}
\end{center}
\vskip 0.35cm
\vbox{
\centerline{
\epsfxsize=2.8 true cm
\epsfbox{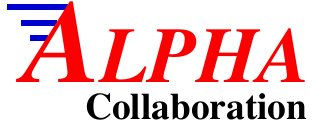}}
}
\vskip 0.1cm
\begin{center}
{
Stefan Schaefer$^{\scriptscriptstyle a}$,
Rainer Sommer$^{\scriptscriptstyle b}$, 
Francesco Virotta$^{\scriptscriptstyle b}$
}
\vskip 0.5cm
{
\vskip 2.0ex
$^{\scriptstyle a}$
        Humboldt Universit\"at zu Berlin, Institut f\"ur Physik, 
	Newtonstr.~15, 12489~Berlin, Germany
\vskip 2.0ex
$^{\scriptstyle b}$
NIC, DESY,
Platanenallee 6, 15738~Zeuthen,  Germany
\vskip 2.0ex
}
\vskip 0.5cm
{\bf Abstract}
\end{center}
\vskip 0.1ex
We study the critical slowing down towards the continuum limit of lattice
QCD simulations with Hybrid Monte Carlo type algorithms.
In particular for the squared topological charge we find it to be very severe
with an effective dynamical critical exponent  of about 5 in pure gauge theory.
We also consider Wilson loops which we can
demonstrate to decouple from the modes which slow down the topological
charge. Quenched observables are studied and a comparison to simulations 
of full QCD is made.
In order to deal with the slow modes in the simulation, we propose a
method to incorporate the information from slow observables
into the error analysis of physical observables and arrive 
at safer error estimates.


\vskip 2.0ex

\centerline{
September 2010}

\vfill
\eject

\end{titlepage}

\section{Introduction}
In all Monte-Carlo methods, the control of statistical and systematic
errors is the main requirement for reliable calculations.
However, this is frequently made difficult by the phenomenon of critical slowing
down, an increase in computational effort while approaching critical points of a 
theory, beyond the naive scaling with the number of points of the system, due 
to an increase of auto-correlation times.
At first sight, this might not
seem a particularly appealing object of study. The auto-correlation times
are not universal quantities, they depend on the particular discretization
of the theory, the algorithms used and the correlation lengths. However, 
in order to control the statistical uncertainties and make certain that 
the simulation is sufficiently ergodic, it is pivotal to ensure that 
all auto-correlations are much shorter than the total run. The danger
one faces in real-world simulations is that there are auto-correlations, which 
are much longer than the total statistics and therefore cannot be detected
from the simulation itself.

Our object of study is lattice QCD, for which 
recent years have witnessed significant progress in the algorithms. 
In particular, simulating light quarks on large volumes has become feasible
on current computers and control over the chiral extrapolation has improved
accordingly~\cite{lat08,lat09}. Consequently, better control over the cut-off effects is the next
target. Approaching the continuum limit means approaching a continuous phase
transition and therefore critical slowing down is to be expected. The question
is how severe it is and whether fine enough lattices can be reached. 
How fine a lattice is needed for sufficient control of the scaling violations
depends again on the quantity to study, the discretization 
and also on the required accuracy. However,
in particular if the physics of charm quarks is to be studied, at least
lattices with a
lattice spacing down to 0.04fm are required for precision physics. 

The severity of the critical slowing down depends on the algorithm and on the 
observable in question.
An observable with notoriously long auto-correlations for virtually all
algorithms used for either pure Yang-Mills theory or QCD is the global topological
charge. It has been studied over the years using link-update algorithms
for pure gauge theory~\cite{DelDebbio:2002xa} and also in QCD with molecular dynamics based 
algorithms~\cite{Alles1996107,PhysRevD.64.054502,Bernard:2003gq,Jung:2010jt}.
However, let us stress that it is not the topological charge itself which is slow.
Slowly moving modes of the transition matrix of the Markov process are just
particularly prominent in this observable and therefore lead to the long
auto-correlations. The same modes also couple to other observables and
also their auto-correlation times are affected. The amount of coupling of
the modes to the different observables is not known a-priori.

This article serves two purposes. First we study the critical slowing down
of various quantities, i.e. the topological charge, Wilson loops and 
hadronic correlation functions as the lattice spacing is varied over
the range used in contemporary simulations. We will observe that among those
only the charge is affected by very severe slowing down.
If one assumes that the picture does not change drastically while going 
from the quenched theory to fully dynamical simulations, one
can use  the scaling laws of the auto-correlation times from this study, to
set minimal requirements for the total simulation time in full QCD.

The second purpose of this paper is the question of how to deal with
the presence of the slow modes in the data analysis.  In particular we 
will propose a procedure to give conservative estimates of the statistical errors
also in the situation where the slow mode contribution cannot be detected 
directly.

In Sec.~\ref{sec:2} we will therefore  give the
basics of the error analysis of Markov Chain Monte Carlo data. This will lay the
ground for the improved error estimates in the presence of very slow modes
of the Monte Carlo evolution. Preparing for the 
numerical results (Sec.~\ref{sec:4}) we list the 
algorithms and observables that we study
in Sec.~\ref{sec:3}.

\section{Error estimation\label{sec:2}}

We consider a Markov chain generated by a transition matrix 
\be
   M(q'\leftarrow q)
\ee
giving the probability for the change from a state $q$ to
a state $q'$. For simplicity we assume a discrete set of states
$q$. The desired ensemble distribution, $P(q)$, is an eigenvector of 
the transition matrix, $\sum_q\,M(q'\leftarrow q) P(q) = P(q')$.
Ensemble averages of observables $O_\alpha(q)$ are 
\be \label{e:mean}
 \langle O_\alpha \rangle = \sum_q\,O_\alpha(q)\, P(q)\,. 
\ee 
In the numerical application, after a suitable thermalization, 
we take a finite number of Monte Carlo steps $N$, yielding states 
$q_1,\ldots,q_N$ and estimate 
\bea
   \langle O_\alpha \rangle &=& \Obar_\alpha \pm \delta  \Obar_\alpha \,,\quad
   \Obar_\alpha = {1\over N} \sum_{i=1}^N\,O_\alpha(q_i)\,.
\eea
The uncertainties $\delta  \Obar_\alpha =\rmO(1/\sqrt{N})$ and
more generally those of functions $F(\langle O \rangle)$ 
are given in terms of the auto-correlation function  
\be
  \Gamma_{\alpha\beta}(t) = \lim_{K\to\infty} {1\over K}
  \sum_{i=1}^K \,[O_\alpha(q_{i+t})-\langle O_\alpha \rangle]\, 
                  [O_\beta(q_i)-\langle O_\beta \rangle]  
  \label{e:gammadef}
\ee
and  have to be estimated from the generated finite sequence 
$q_1,\ldots,q_N$ itself. This
is done by  evaluating the expression in \eq{e:gammadef} for a finite but large $K$. 
For the estimate of the error of $\Gamma$ see \app{sec:ee}.  

The formulae
\begin{align}
  (\delta \overline{F})^2 
  &= {\sigma_F^2 \over N} \, 2\tauint(F)\,,  & \sigma_F^2&=\Gamma_F(0) \,, \\
   \tauint(F) &= \frac12  + \sum_{t=1}^{\infty} \rho_F(t) \,, &
      \quad \rho_F(t) &= {\Gamma_F(t) \over \Gamma_F(0)}\,,\\
   \Gamma_F(t) &= \sum_{\alpha,\beta}\,  F_\alpha\Gamma_{\alpha\beta}(t)F_\beta \,,
   \label{e:gammaf}
\end{align}
are derived by a Taylor expansion of 
$F$ in terms of $\langle O_\alpha \rangle$ \cite{Sokal,Madras,UWerr}. For
complicated functions $F$, the occurring
 derivatives 
$F_\alpha = {\partial F \over \partial \langle O_\alpha \rangle}$ can 
be evaluated numerically  \cite{UWerr}.
 
The integrated auto-correlation time, $\tauint(F)$, characterizes 
the dynamics of the Monte Carlo process relevant for the observable
$F$. It is difficult to determine, since the errors of $\Gamma(t)$ 
remain roughly constant as a function of $t$. Therefore the proposed
estimate of
Madras and Sokal \cite{Madras} and its generalization
for functions of primary observables by Wolff involve a window $W$,
\bea
\tauint(F,W) &=& 
      \frac12 + \sum_{t=1}^{W-1} \rho_F(t) \,.
\eea
The window is chosen to balance the systematic error due to 
truncation,
\be
  R_F(W)  =  \sum_{t=0}^{\infty} \Gamma_F(W+t)
  \label{e:tail}
\ee
with the statistical error. In particular \cite{UWerr} 
advocates  the value of $W$ which minimizes an 
estimate\footnote{The exact formula applied in \cite{UWerr} is
\be
\tau^{-1}_W = \log\left (\frac{1+1/(2\tauint(F,W))}
                            {1-1/(2\tauint(F,W))}\right)\Big/S\,.
    \label{e:UWerrW1}
\ee
}
\be
E(W)=  e^{- W/\tau_{W}} + 2\sqrt{W/N}
\quad \text{where}\quad \tau_W \approx S\, \tauint(F,W)
\, 
    \label{e:UWerrW}
\ee
for the sum of  systematic and  statistical relative error of $\tauint$.
$S$ is a parameter, which by default is set to 1.5, and
has to be adjusted by hand if 
other time scales, much larger than $\tauint$ are relevant.
In other words, a proper choice of $S$ requires an inspection 
of the particular shape of the 
auto-correlation function.

We note that this criterion estimates the time scale for
contributions  to $\tauint(F)$ from $t\geq W$ by $\tauint(F,W)$ itself. 
However, when the lattice spacing becomes small, the time scale which is 
relevant for the tails of auto-correlation functions can become
significantly different from $\tauint(F)$ in lattice gauge theory
simulations. We will see examples of this in \sect{sec:4}. 
Indeed, it can be shown that 
$
   |\Gamma_F(t)|  \leq  \mathrm{const.}\,\mathrm{e}^{-t/\tauexp}
$
for any Markov chain~\cite{LH:martin}. 
An elegant proof is given in the cited reference. 

It is usually assumed that the above bound is realized at large $t$ vs.
\be
    \Gamma_F(t) \simas{t\to\infty} A_F\,\mathrm{e}^{-t/\tauexp} 
    \label{e:slowmode}
\ee
up to terms with a faster exponential decay.  Indeed for 
algorithms which satisfy the detailed balance condition,
\be
   \label{e:detbal}
   M(q'\leftarrow q)\,P(q) = M(q \leftarrow q')\, P(q')\,,
\ee
amongst them 
most versions of the Hybrid Monte Carlo (HMC) algorithm \cite{hmc},
\eq{e:slowmode} can be proven. 
We turn to a brief discussion of auto-correlation functions
in this more restricted class.

\subsection{Algorithms with detailed balance\label{sec:db}}

When \eq{e:detbal} is satisfied,
it is convenient to introduce the symmetric matrix
\be
  T(q,q') = [P(q')]^{-1/2}\,M(q'\leftarrow q) \,[P(q)]^{1/2}\,,
\ee
which has real eigenvalues $\lambda_n, \;n\geq0$, with 
$\lambda_0=1$ and $|\lambda_n|<1$ for $n\geq1$, assuming
an ergodic algorithm. We order the eigenvalues as  
$\lambda_n\leq\lambda_{n-1}$. There is a complete set of 
eigenfunctions $\chi_n(q)$ with  $\chi_0(q)=[P(q)]^{1/2}$. 
Starting from the representation 
\be
  \Gamma_{\alpha\beta}(t) = 
    [O_\beta(q')-\langle O_\beta \rangle]\,  M^t(q'\leftarrow q)\, 
    [O_\alpha(q)-\langle O_\alpha \rangle]\, P(q) 
\ee
with 
$M^{n+1}(q'\leftarrow q)=\sum_{q''}M(q'\leftarrow q'')M^{n}(q''\leftarrow q)$,
we then have
\bea
  \Gamma_{F}(t) &=& \sum_{\alpha,\beta}\,F_\alpha F_\beta \, \sum_{q,q'}
    [O_\alpha(q)-\langle O_\alpha \rangle]\,  
    [P(q)]^{1/2}\, T^t(q,q')\,[P(q')]^{1/2}\, 
    [O_\beta(q')-\langle O_\beta \rangle] 
  \nonumber \\
  &=&  \sum_{n\geq1} (\lambda_n)^t\; [\eta_n(F)]^2 
  \label{e:spect}
\eea
in terms of the ``matrix elements'' 
\be
  \eta_n(F) = \sum_{\alpha}F_\alpha\,  \sum_{q} \chi_n(q)[P(q)]^{1/2} \,
              [O_\alpha(q) - \langle O_\alpha\rangle ]\,.
\ee
We recognize  \eq{e:slowmode} with $A_F=[\eta_1(F)]^2$ and 
$\tauexp=-1/\log(\lambda_1)$ provided $\lambda_1>0$.
In general all eigenmodes of the matrix $T$ contribute to the above sum over
$n$. 

However, exact symmetries may entail selection rules with $\eta_n(F)$ vanishing
for some $n$.  As an example let us consider a parity symmetry
$q\to q'=S(q)$ with 
$P(S(q))=P(q)$ and $S(S(q))=q$.  It is a symmetry of the algorithm
if  
\be
  T(S(q'),S(q)) = T(q',q)\,. 
\ee
With respect to the action of $S$, 
the eigenfunctions $\chi_n(q)$ of $T$ can then be divided into even ones,
$\chi_{n_+}(S(q)) = \chi_{n_+}(q)$ and odd ones,
$\chi_{n_-}(S(q)) = -\chi_{n_-}(q)$. Observables are then also split
into even ($s=1$) and odd ($s=-1$), 
$F_s(O(S(q)))=s\,F_s(O(q))$ and have an auto-correlation function
\be
  \Gamma_{F_s}(t) =  \sum_{n_s\geq1} (\lambda_{n_s})^t\; [\eta_{n_s}(F_s)]^2 
\ee
with only even or odd contributions. Since the ensemble average of odd
observables vanishes, one can restrict the attention to $s=+1$.

Most versions of the HMC algorithm for QCD are invariant under ordinary 
parity,
which means that it suffices to look at parity invariant observables
to search for the relevant slowest mode. For our QCD studies we 
therefore consider $Q^2$ instead of the parity odd topological charge $Q$.

We are now in the position to discuss improved error estimates, 
namely estimates which aim at giving a realistic and/or 
conservative estimate of the tail contribution \eq{e:tail}
to the error of $F$ also in the situation when $\tauexp$ is significantly 
larger than $\tauint(F)$.

\subsection{Improved error estimates\label{sec:iee}}

Remaining with algorithms which satisfy detailed balance,
we can start from \eq{e:spect}. For $n\geq1$ we then have
$|\lambda_n|<1$ and
 $\sum_{t=0}^{\infty}\,  (\lambda_n)^t = 1/(1-\lambda_n)>0$ and
furthermore $ 1/(1-\lambda_n) \leq  1/(1-\lambda_1)$.
This yields bounds
\bea
   R_F(W) &\leq& {1 \over 1-\lambda_1}
                   \sum_{n\geq1}  (\lambda_n)^{W}[\eta_{n}(F)]^2
              =  {1 \over 1-\lambda_1} \Gamma_F(W) 
              = \tauexp\Gamma_F(W)\,(1 + \rmO(1/\tauexp))\,,
   \nonumber \\[-1ex]
   \label{e:uppbound}\\ 
   R_F(W) &\geq& {1 \over 1-\lambda_1}  (\lambda_1)^{W}\,
            [\eta_{1}(F)]^2 = \tauexp\,{\rm e}^{-W/\tauexp}\,(1 + \rmO(1/\tauexp))
   \label{e:lowbound}
\eea
for even $W$. They translate into bounds on $\tauint(F)$.

As long as the configuration space is large,
we expect these bounds to hold quite generically, also for
algorithms which do not satisfy detailed balance. Certainly Monte Carlo (MC)
experiments that we have seen so far are in agreement with
such a behaviour. 

Let us now {\em assume} that we are in a situation where the following
is true
\begin{itemize}
\item[1.] There is some knowledge about $\tauexp$ from previous 
   MC runs or an extrapolation from other parameters of the
   simulated theory.
\item[2.] The considered MC run is still long compared to $\tauexp$ itself,
   \be  
      N \gg \tauexp \,,
      \label{e:condition2}
   \ee
   but not so long that one can just sum up the auto-correlation function
   with a window $W\sim\tauexp$.
 \item[3.] We are interested in an error estimate which safely includes 
   the contribution represented by the slow mode corresponding to
   $\tauexp$ or slow modes $n$ with $\lambda_n\approx \lambda_1$.
\end{itemize}
We propose to choose a window $\Wl$, according to the criterion of \eq{e:UWerrW} and explained in \cite{UWerr}, with
the parameter $S$ set to its default value of 1.5 and the associated

\be \label{e:tauintl}
  \tauintl(F) = \tauint(F,\Wl)
\ee
as well as a second window  $\Wu$ where the 
auto-correlation function is still significant by, e.g.,
three standard deviations and add an estimate of the tail
giving 
\be
  \label{e:tauintu}
  \tauintu(F) = \tauint(F,\Wu)+\tauexp \rho_F(\Wu)\,.
\ee
In cases where $\rho_F$ falls very quickly and is compatible with zero
at short time $t=W_0$, e.g. $W_0=5$, we replace this estimate by
\be
  \label{e:tauintup}
  \tauintu(F) = \tauint(F,W_0)+2\tauexp \delta[\rho_F(W_0)]
   \quad\text{for}\quad \delta[\rho_F(W_0)] >  \rho_F(W_0) \,,
\ee
where $\delta[\rho]$ is the estimated error of $\rho$.
When one is interested in $\tauint(F)$ itself, e.g. for the investigation of
algorithms, one should choose an interval covering  $\tauintl(F)$ and
$ \tauintu(F)$ together with their statistical errors. If on the other hand
one just wants a safe estimate of the error of the observable
we propose to choose $\tauintu(F)$.
 
\begin{figure}[htb!]
\begin{center}
\includegraphics{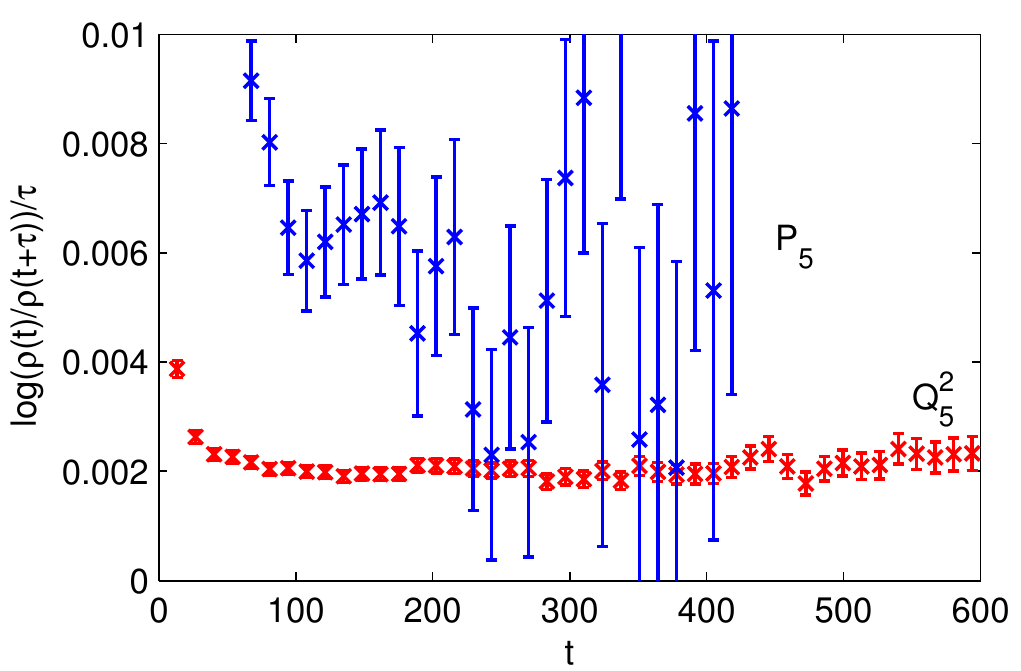}
\end{center}
\caption{\label{f:effmass}Effective mass plot 
$\frac{1}{\tau}\log(\rho_F(t)/\rho_F(t+\tau))$ for two observables
$F$ in run C3d. Here $\tau=6.75$ is the 
spacing between consecutive measurements.}
\end{figure}

An additional issue is that in the presumed situation,
it is also of interest to estimate how severely an observable $F$ 
is affected by the slow mode(s). The ratio $\tauintu(F)/\tauexp$
is a possible measure, but to quantify this more precisely, 
it is better to try to isolate the contribution of the slowest mode. 
The corresponding normalized amplitude is
\be 
   C_F= {A_F\over \Gamma_F(0)} = \lim_{t\to\infty}
 \rho_F(t) \mathrm{e}^{t/\tauexp} \,.
   \label{e:CF}
\ee
One may object immediately that it is very difficult, if not impossible, 
to estimate $\tauexp$, which is needed in the above formulae. 
In \fig{f:effmass} we therefore just show one numerical result already 
at this point: the ``effective mass plot'' from auto-correlation functions
of a few observables.
Details of the numerical simulation are described only later in \sect{sec:pg},
however, it is clear from the figure that considering several observables
can help for getting a handle on the slow modes.
Of course the statistics has to be large enough, but as an empirical
observation, an early onset of the plateau in  $\log(\rho(t)/\rho(t+\tau))$ 
is beneficial  when $\tauexp$ is large.
Furthermore, the whole proposal relies on the fact the slowest mode,
and with it $\tauexp$, can actually be identified. Absolute certainty 
on this is virtually impossible to achieve, however, by looking at a large number
of operators, at least a significant portion of the relevant space
can be covered. Also in case there is an even slower mode than the
one identified, the proposed method does provide a more
conservative estimate of the contributions up to this threshold,
and can therefore improve the analysis.

\subsection{Decoupling and dynamical correlation coefficient 
                   \label{sec:ddcc}} 

Since $\tauexp$ enters in the exponent in \eq{e:CF},
this representation is useful if $\tauexp$ is already 
known rather precisely -- a rare luxury. A more practical
representation replaces $\tauexp$ by an effective one.
To this end, we take observables $O_\beta$ which couple relatively 
strongly to the slow MC mode. For QCD  possible choices are 
the square of the topological charge $O_\beta=Q^2_\alpha$ or the smeared
plaquette $O_\beta=P_\alpha$ with  $\alpha$ labelling different
smearing levels, see \sect{sec:obs} for details. 
\footnote{We
remind the reader that in QCD with parity conserved, 
the whole discussion is to be 
restricted to parity even observables.}  
We can use
\be \label{e:tauexpeff}
 \tauexp^{\mathrm{eff}}(t) = \frac{t}{2
 \log\left\{\mathrm{Max}_\beta{\rho_{\beta}(t/2) \over \rho_{\beta}(t)}\right\}}\,,
\ee
but clearly other choices are possible. The effective
coefficient
\be 
   C_F^\mathrm{eff}(t)= \rho_F(t) \mathrm{e}^{t/\tauexp^{\mathrm{eff}}(t)} \,.
   \label{e:CF1}
\ee
then suggests itself.
When detailed balance is guaranteed, a further effective
estimator is  
\be
   \widetilde{C}_F^\mathrm{eff}(t)= 
 {[\Gamma_{FG}(t)]^2 \over \Gamma_{F}(0)\Gamma_{G}(t)}
 \mathrm{e}^{t/\tauexp^{\mathrm{eff}}(t)}
   \label{e:CF2}   
\ee
where $\Gamma_{FG}(t) = \sum_{\alpha,\beta}F_\alpha\Gamma_{\alpha\beta}(t)G_\beta$ 
and we have assumed that $G$ is an observable with a strong
coupling to the slow mode. In other words $C_{G}$ is large. 
This representation will be valid
(at large $t$) if $\lambda_1$ is an isolated eigenvalue and
in practice if indeed the critical slowing down is dominated
by the single mode $n=1$. It simply follows from the mode
decomposition $\Gamma_{FG}(t)= \sum_{n\geq1} (\lambda_n)^t\; \eta_n(F)\eta_n(G)$.

Clearly \eq{e:CF1} is more generic
and even expected to be useful when detailed balance is not 
satisfied, but the advantage of \eq{e:CF2} is that it can possibly
be used at much larger $t$, showing smaller statistical errors in that region. 

We can now define what we mean by decoupling of an
observable from the slow mode $n=1$: in practice 
it means $C_F \ll 1$ while in terms of critical
slowing down, it should be defined as a significant
decrease of $C_F$ as the correlation length and $\tauexp$
grow, e.g. $C_F \sim (\mathrm{correlation\; length})^{-\gamma}$ with
some positive $\gamma$. In MC runs this decoupling is expected to be
visible in the behaviour of $\widetilde{C}_F^\mathrm{eff}(t)$
at moderate time $t$. Given the inherent problems in seeing
asymptotic behaviour in numerical simulations,
it is useful to go further and define a time scale $\tau_*$
through 
\be
  \tauexp^{\mathrm{eff}}(r\,\tau_*) = \tau_*
   \label{e:taustar}
\ee
and
\be 
   C_F^{*}(r)=  C_F^\mathrm{eff}(r\tau_*)\,.
   \label{e:CFstar}
\ee
In the same way, $C_F^\mathrm{eff}$ may be replaced by $\widetilde{C}_F^\mathrm{eff}$.
Using $\tau_*$ is similar in spirit to the original Sokal proposal for fixing the window of
summation for the $\tauint$ by  the point at which the summation window $W$ exceeds a
multiple of $\tauint(W)$.
A choice of $r$ significantly smaller than one is necessary when the
overall statistics is moderate. We emphasize again our condition \eq{e:condition2}, however. 
The advantage of \eq{e:CFstar} is that we do not have to consider 
asymptotically large $t$ with their associated systematics. Decoupling
can be studied at a fixed (not unreasonably small) value of $r$. 
If $C_F^{*}(r)$ shows decoupling it will usually also be the case in $C_F$.

\subsubsection{Relation to static correlations \label{sec:sc}} 
In the language used here, the {\em square of} the standard correlation
coefficient of observables $F$ and $G$ is\footnote{Normally one will just consider
primary observables, but the correlation coefficient of arbitrary 
functions $F,G$ is a straightforward generalization.}
\bea
  C^\mathrm{static}_{FG} = {[\Gamma_{FG}(0)]^2 
     \over \Gamma_{F}(0) \,\Gamma_G(0)}\,.
\eea
It is a static property, independent of the algorithm as
only $t=0$ appears. We now notice that {\em if} $G$ ``is'' approximately 
the slow mode, which precisely means
\bea
  |\eta_1(G)| \gg  |\eta_n(G)| \;\;\forall \; n>1 \, ,
\eea
then we have
$ \Gamma_{FG}(t)  \approx \lambda_1^t\, \eta_1(F)\eta_1(G) $
and $ \Gamma_{G}(t)  \approx \lambda_1^t\, [\eta_1(G)]^2 $
and therefore also
\bea
  C_{F} \approx C^\mathrm{static}_{FG}\,.
\eea
We may therefore consider $C_{F}$ to be the dynamical correlation coefficient
between $F$ and $G$.
Generically it will be very different from the static one, emphasizing that
the static correlation of the observable $F$  to a slow one (e.g. the topological
charge in QCD) is not the proper way to discuss the error of $F$. 
Rather the dynamical correlation coefficient $C_{F}$ has to be used.

\section{Algorithms and observables under study\label{sec:3}}
For the numerical investigation of the HMC and DD-HMC, 
we now first give the basic definition of the algorithms and then
of the observables which we choose to investigate.

\subsection{Algorithms \label{sec:algo}}

In hybrid Monte Carlo~\cite{hmc} and related algorithms, the gauge fields
are updated by solving the classical equations of motion associated with the
Hamiltonian 
\be
H=\frac{1}{2}(\Pi,\Pi)+S(U) \ ,
\ee 
where the antihermitian $\Pi_{x,\mu}$ are the momenta  conjugate to
the gauge fields $U_{x,\mu}$. Their scalar product is defined as $(\Pi,\Pi)=-2\sum_{x,\mu}
\mathrm{tr}\Pi_{x,\mu}^2$. With the Monte Carlo time $\tau$, the equations of motion
then read
\be
\frac{\mathrm{d}}{\mathrm{d}\tau} \Pi_{x,\mu} = - \Force_{x,\mu} \ \ \text{and} \ \ 
\frac{\mathrm{d}}{\mathrm{d}\tau} U_{x,\mu} = \Pi_{x,\mu} U_{x,\mu}  \ ,
\ee
where the force $\Force$  fulfills $(\omega,\Force)=\delta_\omega S(U)$ for infinitesimal
variations of the gauge field $\delta_\omega U_{x,\mu} = \omega_{x,\mu}
U_{x,\mu}$. In these definitions, we follow the ones used in the context of the
DD-HMC\cite{algo:L2}. We give them, because they fix the normalization of the
trajectory length $\tau$, which is not unique in the literature. The
conventions of  Ref.~\cite{Gottlieb:1987mq}  used, e.g., in the MILC
code result in a different normalization: a
trajectory of length $\sqrt{2}$ in the conventions above corresponds  to a unit
length trajectory in those of Ref.~\cite{Gottlieb:1987mq}.

The difference between HMC and DD-HMC is that the latter introduces a
decomposition of the lattice into blocks of size $B_0\times B_1 \times B_2
\times B_3$. During each trajectory, only the links are updated, which
have at least one endpoint in the interior. The fraction of  these ``active'' links is given by
\be
R=\frac{\prod_{i=0}^3(B_i-2)+\frac{1}{4}\sum_{i=0}^3\prod_{j\neq i}(B_j-2)}
       {\prod_{i=0}^3B_i} \ .
\ee
Since the active links are treated in exactly the same way as in HMC,
naively, auto-correlation times will be proportional to the inverse of $R$. 
Therefore, we scale them in the following by this ratio, noting that
in pure gauge theory also the cost of the simulation scales accordingly.

At the end of each trajectory, the HMC algorithm has a global Metropolis
acceptance step to correct for the errors in the numerical integration of the
equations of motion. For the DD-HMC in pure gauge theory with the Wilson gauge
action, however, the molecular dynamics evolution of the active links on each
block is independent of the other blocks. We can therefore perform the
Metropolis step for each block individually.\footnote{We thank M.~L\"uscher for
this suggestion.} Compared to the conventional global acceptance step, a given
acceptance rate can be achieved with a significantly larger step size. All our
runs are done at acceptance rates above $90\%$, and in this case, the
block-wise acceptance does not influence the auto-correlation times of the pure
gauge observables within errors.

In order to be ergodic, all links of the lattice have to become active
within some (composite) series of update steps. 
This is achieved by translating the domain decomposition 
relative to the lattice between trajectories. The scheme is described in
detail in Ref.~\cite{algo:L2} and 
alternates random shifts with directed ones, the latter to increase the
efficiency of this step. Because of the  directed shifts, however, the full
algorithm does not obey detailed balance. Even if \eq{e:uppbound}
 can then not be shown mathematically, 
we still expect it to be valid at not too small $t$. In any case 
\eq{e:tauintu} represents a useful estimate of the integrated
auto-correlation time. 
The same reasoning holds true for the factorization behind \eq{e:CF2}.

\subsection{Observables \label{sec:obs}}
We want to study the effect of the  critical slowing down of the (DD)-HMC 
algorithm on observables of interest for physics. We consider
meson two-point functions, Wilson loops and the topological charge,
which we will now define. In order to be more sensitive to the slow modes,
we also computed some observables on smoothed gauge fields.
For this purpose we apply up to five levels of  HYP smearing\cite{HYP}
to the link variables.

The slow evolution is very prominent in the topological charge, for  which we
use  the naive gauge definition
\begin{equation}
\label{eq:Q}
Q_\alpha=\frac{1}{16 \pi^2} a^4 \sum_{x,\mu,\nu} \,  {\rm tr}
    \left [ F_{\mu\nu}^{(\alpha)}(x) \tilde F_{\mu\nu}^{(\alpha)} (x)\right] \ ,
\end{equation}
where the lattice field strength tensor $F^{(\alpha)}_{\mu\nu}(x)$ is constructed from 
the clover leaf representation (see e.g. \cite{impr:pap1} for a definition) 
but from $\alpha$ times HYP smeared
links, where we consider $\alpha\leq5$. We find little difference beyond
the first iteration of smearing. 

As physics oriented observables, we compute $W_1(l_1,l_2)$, the Wilson loops of 
size $l_1\times l_2$ after
one level of HYP smearing, and  the ones without smearing $W(l_1,l_2)$. 
Only the plaquette $P_\alpha = W_\alpha(a,a)$ is also considered 
with higher levels of smearing, $\alpha\leq5$.

In order to study the effects of the slow modes on hadronic observables, we
take as an example the correlators used in the quenched study of the 
$D_s$ meson at  parameters of Ref.~\cite{fds:quenched2}. We compute 
\be
\begin{split}
C_\mathrm{PP}(t)&=a^3\sum_{\vec{x}} \langle P^{rs}(t,\vec{x})  P^{sr}(0,0)\rangle\\
C_\mathrm{AP}(t)&=a^3\sum_{\vec{x}} \langle A_0^{rs}(t,\vec{x})  P^{sr}(0,0)\rangle
\label{eq:corr}
\end{split}
\ee
with the pseudo-scalar density  $P^{rs}=\bar r\gamma_5 s$ and the 
time component of the axial-vector current $A_0^{rs}=\bar r\gamma_0\gamma_5 s$.
These are estimated on each configuration using the one-end method\cite{Sommer:1994gg,Foster:1998vw} 
with 5 stochastic $U(1)$ sources per configuration.
Interesting observables are the effective meson mass $m_\mathrm{eff}$, which is 
defined through
\be
\frac{C_\mathrm{PP}(t+a)}{C_\mathrm{PP}(t-a)}
=\frac{\cosh((t+a-T/2)m_\mathrm{eff}(t))}{\cosh((t-a-T/2)m_\mathrm{eff}(t))}
\ee
and the PCAC quark mass ($\partial_tf(t)=\frac1a(f(t+a)-f(t))\,,\;
\partial_t^*f(t)=\frac1a(f(t)-f(t-a))$)
\[
m
=\frac{\frac{1}{2}(\partial_t+\partial_t^*)C_\mathrm{AP}(t)
+a\,c_\mathrm{A}\,\partial_t^*\partial_tC_\mathrm{PP}(t)}{2\,C_\mathrm{PP}(t)} \ .
\]
For both masses, as well as for the decay constant,
\be
 f_\mathrm{PS}(t) = 
\frac{C_\mathrm{AP}(t)}{[C_\mathrm{PP}(t)\, m_\mathrm{eff}(t)]^{1/2}}
\mathrm{e}^{t\; m_\mathrm{eff}(t)/2}\,,
\ee
we average over a suitably chosen plateau in $t$.

\begin{table}[!p]
\small
\begin{tabular}{clcccclllcr}
\hline \hline 
TAG & $\;\;\beta$ & $L/a$ & $T/a$& $a$[fm] & block      & $~~R$ & $\;\tau$ & $\;\Delta \tau$ & A  & stat~ \\
\hline 
A1a & 5.789 & 16 & 16 & 0.140 & $8^4$ & 0.369 & 0.5 & 0.01 & 0.961 & 105280 \\ 
A1b & 5.789 & 16 & 16 & 0.140 & $8^4$ & 0.369 &   1 & 0.01 & 0.971 & 70080 \\ 
A1d & 5.789 & 16 & 16 & 0.140 & $8^4$ & 0.369 &   4 & 0.01 & 0.968 & 141120 \\[1ex]
B0a & 6 & 24 & 24 & 0.093 & HMC & 1 & 0.5 & 0.0077 & 0.931 & 199600 \\
B0b & 6 & 24 & 24 & 0.093 & HMC & 1 & 1  & 0.0077 & 0.954 & 110000 \\
B0c & 6 & 24 & 24 & 0.093 & HMC & 1 & 2  & 0.0077 & 0.943 & 210000 \\
B0d & 6 & 24 & 24 & 0.093 & HMC & 1 & 4  & 0.0077 & 0.946 & 130000 \\
B0e & 6 & 24 & 24 & 0.093 & HMC & 1 & 8  & 0.0077 & 0.945 &  116000\\
B1a &     6 & 24 & 24 & 0.093 & $12^4$ &  0.53 & 0.5 & 0.0077 & 0.932 & 52640 \\ 
B1b &     6 & 24 & 24 & 0.093 & $12^4$ &  0.53 &   1 & 0.0077 & 0.951 & 55520 \\ 
B1c &     6 & 24 & 24 & 0.093 & $12^4$ &  0.53 &   2 & 0.0077 & 0.945 & 61280 \\ 
B1d &     6 & 24 & 24 & 0.093 & $12^4$ &  0.53 &   4 & 0.0077 & 0.945 & 65440 \\ 
B2a &     6 & 24 & 24 & 0.093 & $12^2 \times 6^2$ & 0.363 & 0.5 & 0.0077 & 0.945 & 113800 \\ 
B2b &     6 & 24 & 24 & 0.093 & $12^2 \times 6^2$ & 0.363 &   1 & 0.0077 & 0.958 & 116400 \\ 
B2c &     6 & 24 & 24 & 0.093 & $12^2 \times 6^2$ & 0.363 &   2 & 0.0077 & 0.956 & 119200 \\ 
B2d &     6 & 24 & 24 & 0.093 & $12^2 \times 6^2$ & 0.363 &   4 & 0.0077 & 0.954 & 110400 \\ 
B3a &     6 & 24 & 24 & 0.093 & $6^4$ & 0.247 & 0.5 & 0.0077 & 0.956 & 61000 \\ 
B3b &     6 & 24 & 24 & 0.093 & $6^4$ & 0.247 &   1 & 0.0077 & 0.966 & 128000 \\ 
B3c &     6 & 24 & 24 & 0.093 & $6^4$ & 0.247 &   2 & 0.0077 & 0.963 & 138000 \\ 
B3d &     6 & 24 & 24 & 0.093 & $6^4$ & 0.247 &   4 & 0.0077 & 0.962 & 147000 \\
B4a &     6 & 24 & 24 & 0.093 & $12 \times 6^3$ & 0.3 &   0.5 & 0.019$^*$ & 0.97 & 1008000 \\
B4b &     6 & 24 & 24 & 0.093 & $12 \times 6^3$ & 0.3 &   1   & 0.02$^*$  & 0.97 & 1584000 \\
B4c &     6 & 24 & 24 & 0.093 & $12 \times 6^3$ & 0.3 &   2   & 0.02$^*$  & 0.98 & 780000\\[1ex]
C1d & 6.136 & 32 & 64 & 0.075 & $16 \times 8^3$ & 0.422 &   4 & 0.02$^*$   & 0.946 & 175360 \\
C2b & 6.179 & 32 & 32 & 0.070 & $8^4$ & 0.369 &   1 & 0.0059 & 0.956 & 393000 \\ 
C2d & 6.179 & 32 & 32 & 0.070 & $8^4$ & 0.369 &   4 & 0.0222$^*$ & 0.956 & 1568160 \\ 
C3d & 6.179 & 48 & 48 & 0.070 & $12^4$ &  0.53 &   4 & 0.0182$^*$ & 0.919 & 486560 \\
C4d &   6.2 & 32 & 64 & 0.068 & $16 \times 8^3$ & 0.422 &   4 & 0.0229$^*$ & 0.928 & 684000 \\[1ex]
D1d & 6.475 & 48 & 48 & 0.047 & $12^4$ &  0.53 &   4 & 0.0167$^*$ & 0.927 & 707680 \\ 
\hline \hline 
\end{tabular}
\caption{\label{t:params}Parameters of our runs. We give the bare coupling, the
size of the lattice, the lattice spacing from $r_0=0.5$~fm, the block decompostion
in the DD-HMC, the corresponding fraction of active links $R$, the
trajectory length $\tau$ and the step size of the integration $\Delta \tau$ along
with the acceptance rate $A$ and the total statistics in molecular dynamics units.
Runs with blockwise acceptance step are marked with an asterisk on 
the step size.}
\end{table}

\section{Results\label{sec:4}}
We have performed a considerable number of long simulations
allowing for a study of the dependence of
auto-correlations on several parameters. \Tab{t:params}
presents an overview of the pure gauge
theory simulations; on C1 and C4 also quenched measurements were 
carried out.
Most ensembles are lattices generated
with the Wilson gauge action of constant volume $L^4$ with
$L=2.2\,$fm, where the physical scale comes from $r_0/a$ 
of \cite{pot:intermed}
with a nominal value of $r_0=0.5\,$fm\cite{pot:r0}.
We complement this in \sect{sec:qcd} by a comparison to dynamical $N_f=2$  QCD runs.

\subsection{Pure gauge theory\label{sec:pg}}
Let us start the discussion of the results with the pure gauge ensembles
of the Wilson gauge action at constant physical volume, with main interest
on the dynamical critical slowing down of the topological charge and how
it is reflected in other observables. Since we are in pure gauge theory, 
the Wilson loops will serve as prime reference.

\begin{figure}[htb!]
\begin{center}
\includegraphics{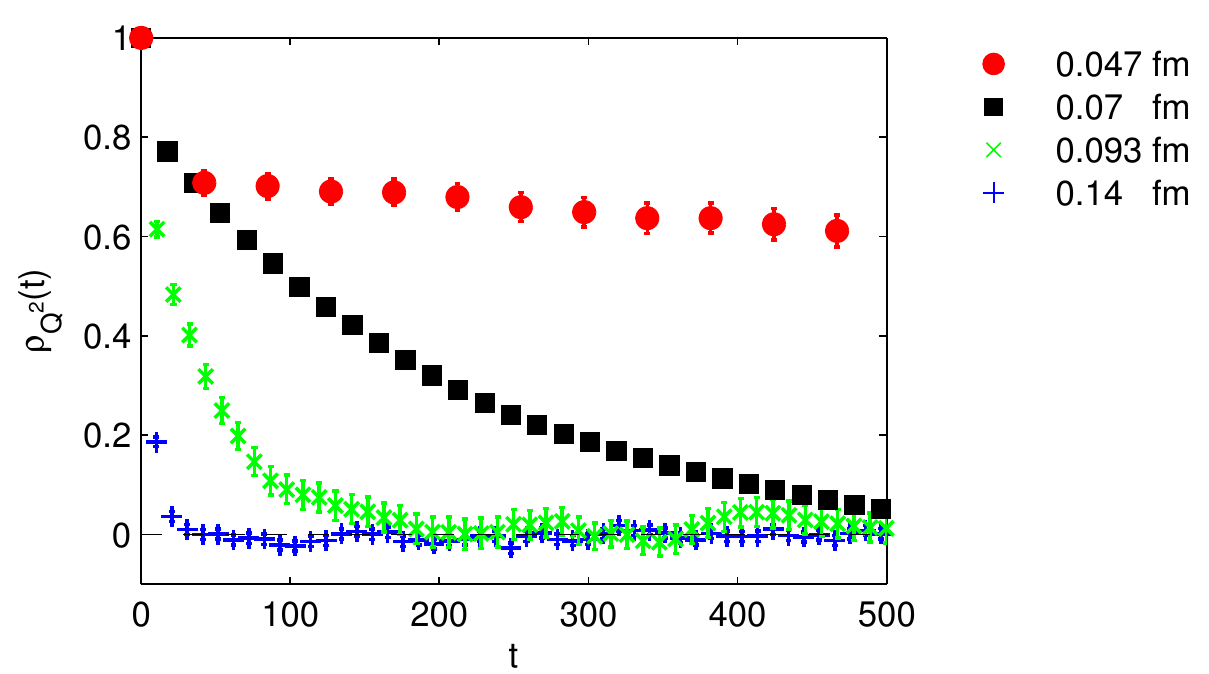}
\end{center}
\caption{\label{f:rhoall}Normalized auto-correlation function of $Q_1^2$ at various
lattice spacings. The Monte Carlo time is given in molecular dynamics units
multiplied by $R$.
}
\end{figure}

\subsubsection{Lattice spacing dependence}

The critical slowing down in the square of the topological charge 
is rather dramatic as demonstrated in \fig{f:rhoall}, where we show
the normalized auto-correlation function for our four lattice spacings,
all with trajectory length $\tau=4$. The Monte Carlo time $t$
is given in molecular dynamics units (MDU) multiplied by $R$. This unit
is applied throughout this paper.

From our data we also determine the integrated auto-correlation times
by using the criterion given in \Eq{e:UWerrW}, where we used the default value
$S=1.5$ unless specified differently. 
Results for 
the plaquette, charge and the square of the charge
as well as auto-correlation times are shown in \Tab{t:charge}. Note that the average
of the charge is compatible with zero in our long runs, 
which is an indication that the determination of uncertainties and
auto-correlations is under control.
We also observe a considerable difference between $\tauint(Q)$ and $\tauint(Q^2)$, 
in line with the arguments given at the end of section~\ref{sec:db}.

\begin{figure}[p!]

\begin{center}
  \includegraphics{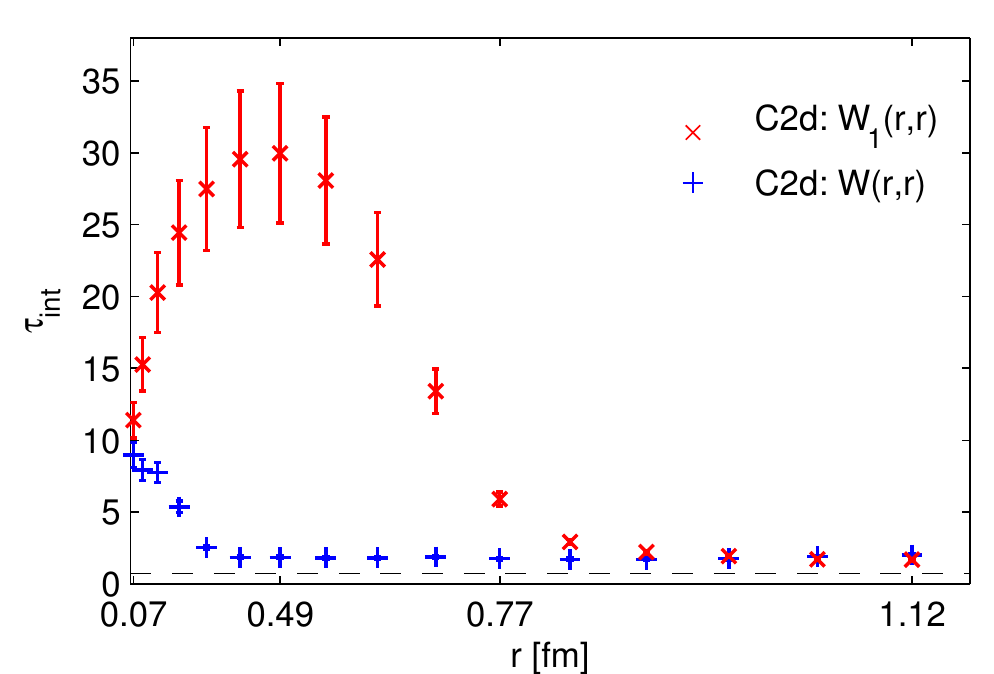}
\end{center}
\caption{\label{f:tauintwloops}Auto-correlation time of square
Wilson loops as a function of their size, for simulation C2d. Smeared 
loops $W_1(r,r)$ 
are marked as $\texttt{x}$, while $\texttt{+}$ symobols show
$W(r,r)$. The line at the bottom shows ${\tau N_mR\over 2}$, where $N_m$ is the number of trajectories of length $\tau$ between two consecutives measurements.}

\end{figure}

\begin{table}[p!]
\begin{center}
{\small
\begin{tabular}[hbtp]{crrrrrrr}
\hline\hline
TAG & $\langle P\rangle \quad$ & $\tauint(P_5)$ & $\langle Q_5\rangle $ & $\tauint(Q_5)$ & $\langle Q^2_5\rangle $ & $\tauint(Q^2_5)$ & $\tau_*$\\
\hline
A1a & 1.697388(57) & 63(15) & -0.05(17) & 32(\p\p6) & 16.8(8) & 18(\p3) & 59(\p7) \\ 
A1b & 1.697473(56) & 32(\p7) & 0.19(15) & 19(\p\p3) & 16.0(5) & 8(\p1) & 34(\p3) \\ 
A1d & 1.697509(55) & 30(\p5) & 0.11(9) & 13(\p\p1) & 16.8(4) & 5.9(5) & 25(\p2) \\[1ex] 
B0a & 1.781044(\p6) & 104(16) & -0.68(22) & 278(\p62) & 18.3(9) & 128(21) & 145(12) \\ 
B0b & 1.781033(\p7) & 56(\p8) & -0.23(20) & 118(\p24) & 19.4(1.0) & 71(12) & 62(\p4) \\ 
B0c & 1.781050(\p6) & 38(\p4) & -0.07(14) & 107(\p16) & 19.3(6) & 51(\p5) & 49(\p3) \\ 
B0d & 1.781049(\p9) & 32(\p3) & -0.14(16) & 87(\p14) & 18.7(6) & 33(\p4) & 44(\p2) \\ 
B0e & 1.781053(14)  & 27(\p3) & 0.02(15) & 74(\p12) & 18.7(7) & 38(\p5) & 41(\p3) \\ 
B1a & 1.781053(13) & 109(36) & 0.83(52) & 214(\p88) & 18.5(1.8) & 84(25) & 104(18) \\ 
B1b & 1.781045(13) & 42(10) & 0.52(43) & 148(\p53) & 19.1(1.6) & 49(12) & 68(10) \\ 
B1c & 1.781055(16) & 44(10) & 0.22(33) & 100(\p30) & 18.3(1.4) & 43(\p9) & 53(\p7) \\ 
B1d & 1.781064(21) & 46(10) & -0.12(33) & 110(\p33) & 17.7(1.3) & 43(\p9) & 50(\p7) \\ 
B2a & 1.781032(13) & 120(35) & -0.41(44) & 212(\p76) & 18.6(1.6) & 78(19) & 107(17) \\ 
B2b & 1.781066(10) & 63(14) & -0.53(41) & 186(\p63) & 19.0(1.3) & 51(11) & 75(10) \\ 
B2c & 1.781067(13) & 44(\p9) & 0.13(32) & 111(\p31) & 20.2(1.4) & 51(11) & 61(\p8) \\ 
B2d & 1.781049(17) & 30(\p5) & -0.15(27) & 80(\p21) & 18.2(1.0) & 34(\p6) & 41(\p4) \\ 
B3a & 1.781072(18) & 79(29) & 0.88(72) & 233(118) & 17.5(2.1) & 58(19)  & 84(17) \\ 
B3b & 1.781057(12) & 48(11) & -0.10(39) & 132(\p44) & 17.8(1.5) & 61(15) & 71(11) \\ 
B3c & 1.781039(17) & 54(13) & 0.05(41) & 156(\p54) & 18.2(1.3) & 45(10) & 57(\p9) \\ 
B3d & 1.781033(21) & 57(13) & -0.04(32) & 105(\p31) & 17.7(1.3) & 48(11) & 51(\p6) \\ 
B4a & 1.781052(\p5) & 97(12) & -0.30(18) & 243(\p43) & 19.2(7) & 114(15) & 135(\p9) \\ 
B4b & 1.781049(\p4) & 62(\p5) & -0.09(10) & 131(\p15) & 18.2(4) & 56(\p4) & 70(\p3) \\ 
B4c & 1.781055(\p6) & 37(\p3) & -0.24(11) & 77(\p10) & 19.5(6) & 52(\p5) & 53(\p2) \\[1ex] 
C1d & 1.822828(\p4) & 42(\p7) & 0.69(61) & 281(\p91) & 49.6(4.1) & 149(38) & 140(18) \\
C2b & 1.835106(\p2) & 80(12) & -0.47(38) & 574(189) & 18.1(1.6) & 248(60) & 376(50) \\ 
C2d & 1.835106(\p7) & 42(\p7) & -1.16(49) & 428(176) & 17.2(1.8) & 178(54) & 159(23) \\ 
C4d & 1.840897(\p2) & 40(\p3) & -0.03(34) & 503(122) & 32.9(1.8) & 217(38) & 249(20) \\[1ex] 
D1d & 1.909347(\p2) & 55(\p5) & 0.90(65) & 4430(2079) & 18.8(2.7) & 2453(959) & 2625(563) \\ 
\hline\hline
\end{tabular}
}
\end{center}

\caption{\label{t:charge}The average plaquette, the topological charge and its
  square along 
with  auto-correlation times (computed with $S=3$) of the smeared plaquette and the (squared) charge 
for our ensembles described in \tab{t:params}. The last
column gives the exponential auto-correlation time as defined in \Eq{e:taustar}.}
\end{table}

\begin{figure}[p!]
\begin{center}
\includegraphics{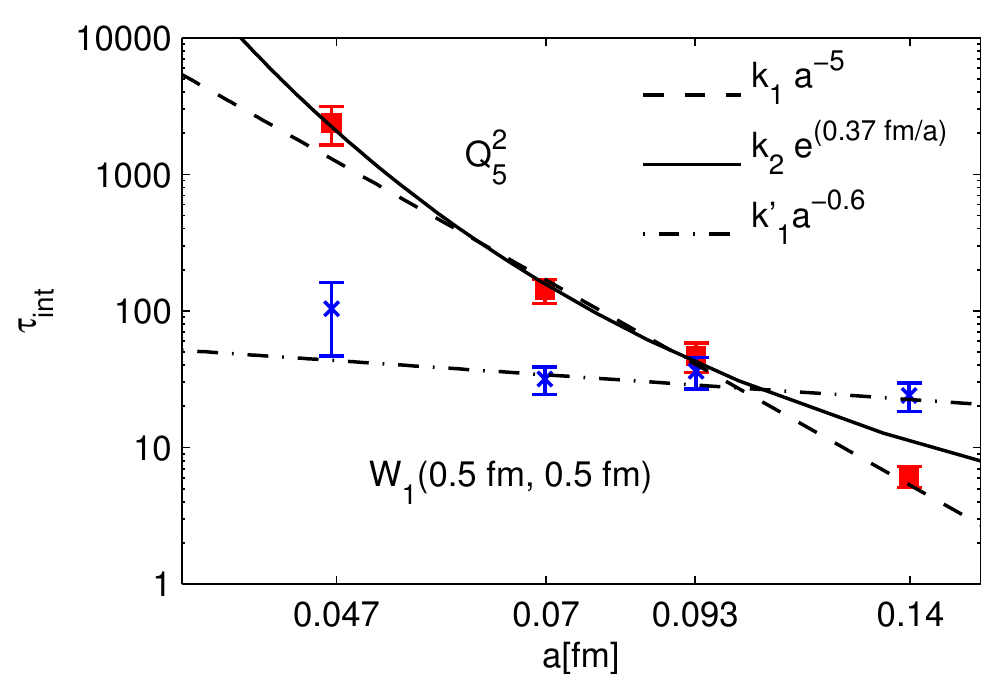}
\end{center}
\caption{\label{fig:tauintvsa}Auto-correlation time of $Q_5^2$
and the $(0.5~{\rm fm})\times(0.5~{\rm fm})$ square Wilson loop
as a function of the lattice spacing using the DD-HMC algorithm. 
For $Q_5^2$ the two curves are fits through the last three points of the two
ans\"atze of \Eq{e:fit}. For the Wilson loop only the fit to the power law has 
been performed, through all points.
}
\end{figure}

The main result of this section is shown in Fig.~\ref{fig:tauintvsa}, where we
give the auto-correlation times of $Q_5^2$ and of $W_1(0.5~{\rm fm},0.5~{\rm fm})$ 
as a function of the lattice spacing. This Wilson loop is chosen since
it is roughly at this size that we find the longest auto-correlation times, see 
\fig{f:tauintwloops}; Creutz ratios behave very similarly. 
The observed maximum of $\tauint$ is surprising at first
sight, but large Wilson loops are dominated by strong ultraviolet (UV) 
fluctuations and therefore have a large variance $\Gamma(0)$ compared 
to their expectation value. In \sect{sec:qa}
we will consider other long distance observables with a smaller variance.

We compare two ans\"atze to describe the behaviour of the
auto-correlation times, 
\be\label{e:fit}
 \tauint(F) = k_1\,(a/r_0)^z \qquad \text{and} 
 \qquad \tauint(F) = k_2\,\exp(\alpha/a)
\ee
where the first is the standard behavior in the vicinity of a continuous phase
transition, whereas the exponential form was advocated in the context
of the $CP^{(N-1)}$ model in \Ref{DelDebbio:2004xh}; we use it only for the
topological charge.
Even our high statistics data is not precise enough to accurately 
determine an effective critical exponent. However, with the power law, we get
$z \approx 5$ for $Q_5^2$, a very severe critical slowing down. 
The data is also not good enough to distinguish it from
the exponential form, for which we find $\alpha \approx 0.4$fm.
The Wilson loop, on the other hand, follows a power law with
$z\approx 0.6$  within our range of data, which is a 
surprisingly mild behavior. This already demonstrates the decoupling
discussed in \sect{sec:2}. The Wilson loops decouple
from the slow modes which couple strongly
to the square of the charge. We will come back 
to this subject below.
The exponent for the Wilson loops is compatible with the $z=1$
for HMC in  free field theory\cite{Kennedy:2000ju}.

\begin{figure}[htb!]
\begin{center}
\includegraphics{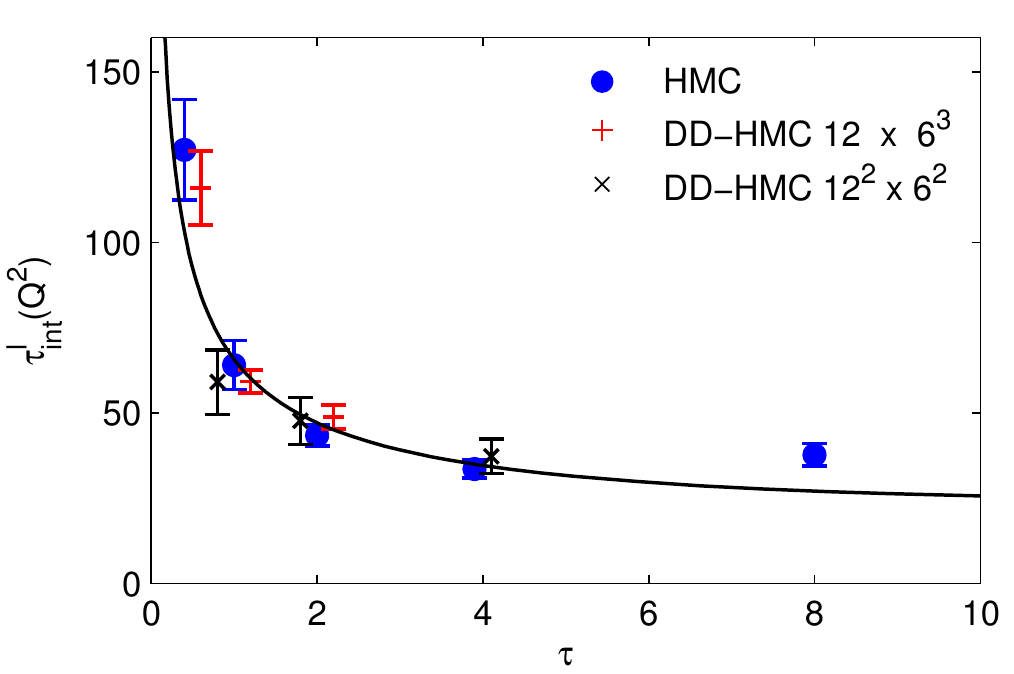}
\end{center}

\caption{\label{fig:tauintvstau}Auto-correlation time of $Q_5^2$
in units of molecular dynamics time scaled by $R$
as a function of the trajectory length for the $24^4$ lattices
at $\beta=6.0$. We show the data for two block decompositions
in the DD-HMC as well as data for HMC simulations. The curve is a fit through all points to the functional form $c_1/\sqrt{\tau}+\tau/2$.
}
\end{figure}

\subsubsection{Dependence on trajectory length and block size}
This brings us to the discussion of the various parameters, on 
which this picture might depend:  the 
trajectory length, the block decomposition and the physical volume.
The dependence of $\tauint(Q_5^2)$ on the trajectory length is visualized 
in \fig{fig:tauintvstau} for 
the $a\approx0.1$fm lattices.
It demonstrates that longer trajectories can lead
to shorter auto-correlation times in units of molecular dynamics
time, which takes into account the additional effort needed for the
longer trajectories. 
That longer trajectories can improve
the performance of the algorithm has been part of the original motivation
for the Hybrid Molecular Dynamics\cite{Gottlieb:1987mq}, 
and has since been demonstrated, e.g., in \Ref{algo:trajlength}.
In free field theory it is known that the optimal trajectory length
depends on the observable and typically increases
when the correlation length increases~\cite{Kennedy:2000ju}.
As long as the system is in a regime with $\tauint\gg \tau$,
one can argue that the momentum refreshment at the 
beginning of each trajectory initiates a random change of 
direction in the otherwise directed walk. One then 
expects longer trajectories to decrease $\tauint$ 
proportional to $1/\sqrt{\tau}$, but at most down to the smallest
possible value of $\tauint=\tau/2$, which means $\tauint=1/2$ in units
of complete updates. 
This simple model describes the gross features of
our data reasonably well. Also
on the finer C lattices, given in \tab{t:charge}, the corresponding 
improvement can be observed.

The data of the figure are also collected in \tab{t:charge} together with
those from different block decompositions and also from the HMC algorithm.
We observe that the blocks do not have a measureable impact on the
auto-correlation times beyond the simple rescaling with the 
active link ratio $R$. Of course, the blocks have to have
a reasonable minimal size. Our smallest blocks are still
at least 0.5fm across, which is around the pure gauge theory 
correlation length defined from the string tension. 

\subsubsection{Dependence on volume and discretization}

\begin{figure}[htb!]
\begin{center}
\includegraphics{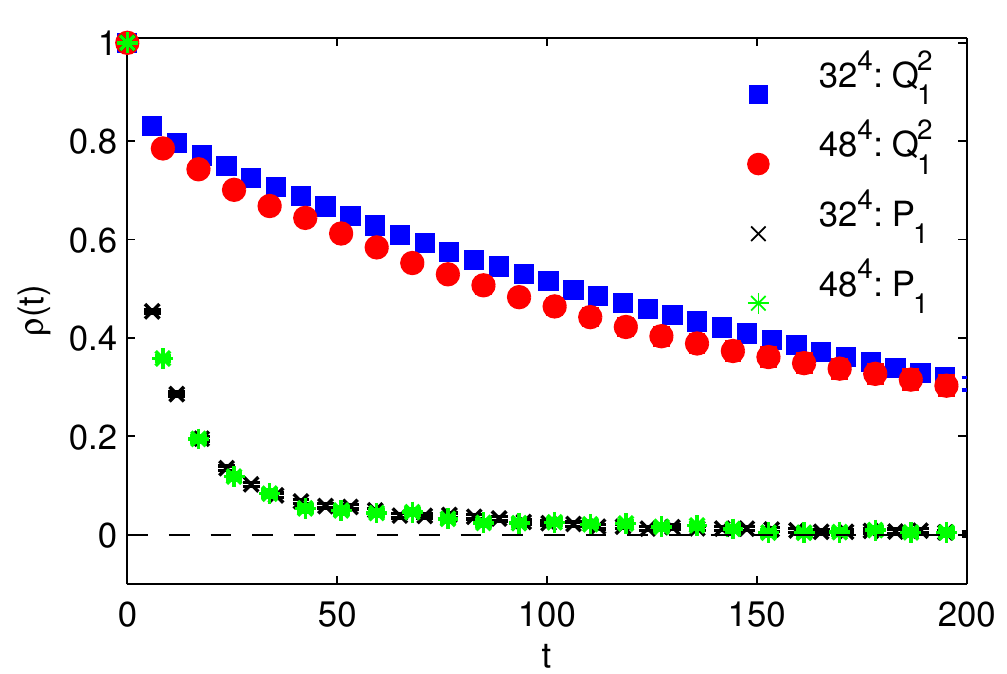}
\end{center}

\caption{\label{fig:volume}Auto-correlation time of $Q_1^2$ and the smeared plaquette $P_1$
at $\beta=6.18$ on a $32^4$ and a $48^4$ lattice.
}
\end{figure}

Most of our ensembles have a constant physical volume with $L=2.2$fm,
for which finite size effects of typical equilibrium expectation
values are known to be small 
in the pure gauge theory. In order to check for a potential $L$-dependence
of auto-correlations, we also
generated an $L=3.3$~fm ensemble at $\beta=6.18$. \Fig{fig:volume} 
demonstrates that no significant volume dependence is present --
neither for the smeared plaquette $P_1$ nor for the squared charge $Q_1^2$.

The emerging picture might also depend on
the particular discretization used. So far, all results were for the Wilson
gauge action. Therefore, we also generated an ensemble with the Iwasaki action
with $a=0.09$fm, with the same volume and simulation  parameters, using the HMC
algorithm in both cases.
We observe a drastically larger auto-correlation time for the topological
charge,
\bea
  \tauint(Q_5^2)=\phantom{0}34(\phantom{0}4) &\quad& \text{for Wilson gauge action},
  \\
  \tauint(Q_5^2)=220(50) &\quad&   \text{for Iwasaki gauge action}
\eea 
However, this is not
replicated in other observables, both the plaquette and the smeared
plaquette having roughly the same auto-correlation times for
the two actions.

\subsection{The charge in subvolumes\label{sec:subvolumes}}

Ultimately one needs to find an algorithm with smaller 
auto-correlations. For this purpose it is 
important to understand more about how the HMC moves
the gauge fields through configuration space. Of course
this is a difficult problem, as we need to reformulate 
it in terms of specific (gauge invariant) observables. 

\begin{figure}[htb!]
\begin{center}
\includegraphics{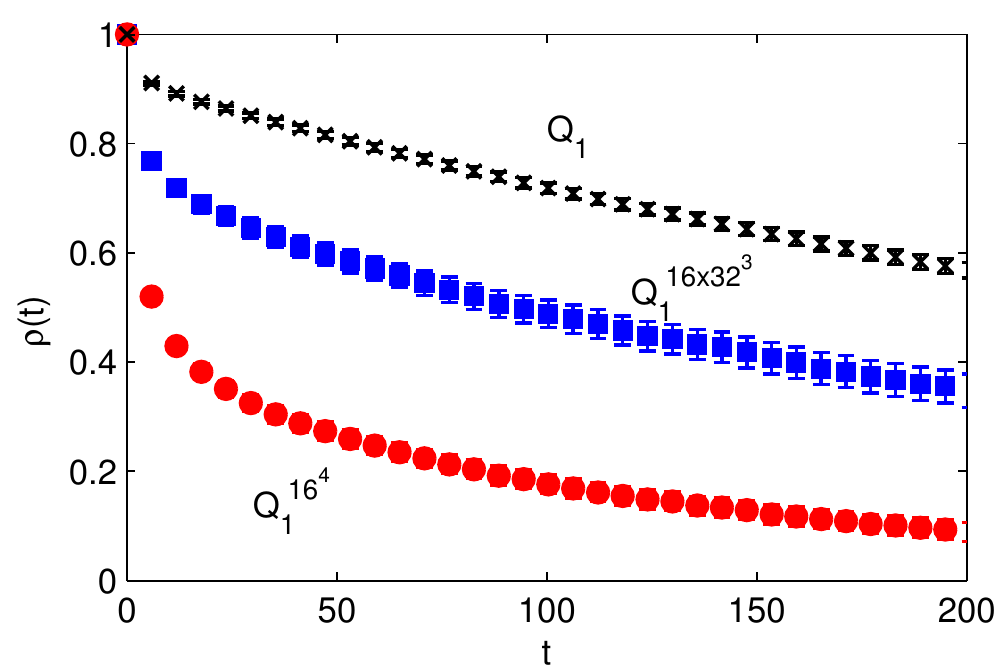}
\end{center}
\caption{Auto-correlation function of $Q_1^{\cal R}$, with ${\cal R}$ being
the full lattice, half the lattice (cut in one dimension)
and a 16th of the lattice, cut in half in all dimensions. 
We used a sequence of 320000~MDU
of the C2d run.\label{fig:subvolume}}
\end{figure}

An interesting such question is whether topological
charge is being moved from some space-time volume to another
one more quickly than the total charge is changing. 
This can be looked at by restricting the sum in \eq{eq:Q}
to a region ${\cal R}$, 
computing the charge inside that region
\begin{equation}
\label{eq:QR}
Q_\alpha^{\cal R}=\frac{1}{16 \pi^2} a^4 \sum_{x \in {\cal R}}
    \sum_{\mu,\nu} \,  
    {\rm tr}
    \left [ F_{\mu\nu}^{(\alpha)}(x) \tilde F_{\mu\nu}^{(\alpha)} (x)\right] \ .
\end{equation}
Its MC history will show whether charge has flown in or out of the 
region. More quantitatively we can directly look at
the auto-correlation function of 
$Q_\alpha^{\cal R}$ as shown in
\fig{fig:subvolume}. The subvolume charge does decorrelate significantly
faster than the total charge, but there is still a quite significant 
coupling to the slow mode remaining. The decoupling coefficient $C^*$ 
is around $0.7$ for the $16\times32$ sublattice and about $C^*=0.15$ for the
$16^4$ subvolume. The latter is a significant suppression.

\subsection{Quenched approximation\label{sec:qa}}

\begin{figure}[htb!]
\begin{center}
\includegraphics{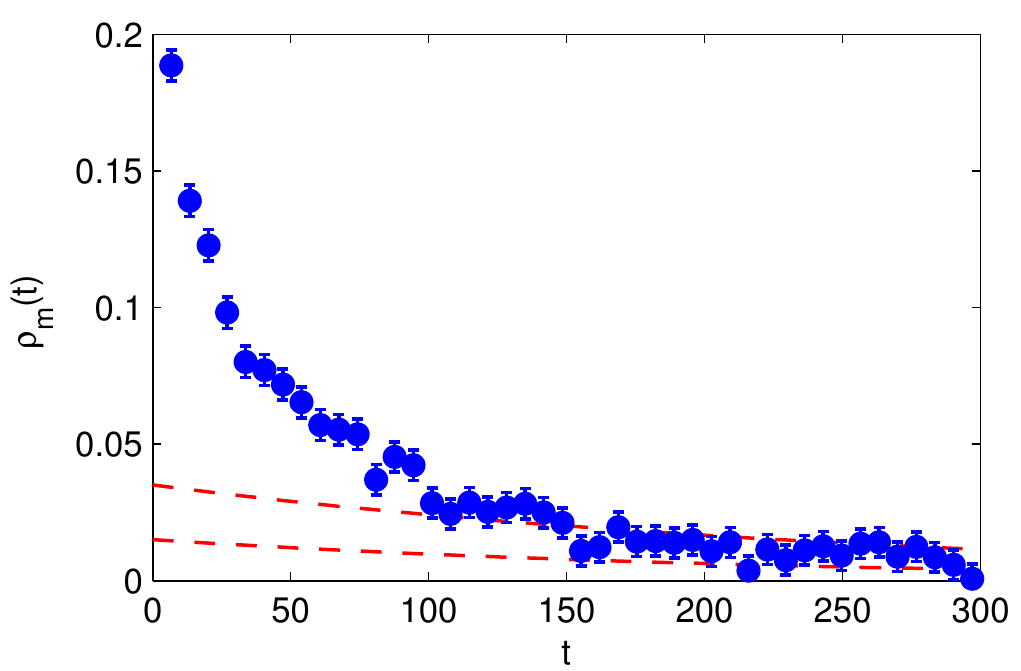}
\end{center}
\caption{
  Auto-correlation function of the mass of the $c \,\bar c'$ pseudo-scalar meson
  with $m_c=m_{c'}=m_\mathrm{charm}$.
  The meson mass is obtained from a plateau average. The two dashed lines show the upper/lower bound region of the tail contribution to the normalized auto-correlation function, given by $C^*_F\;e^{-t/\tau_*}$.
  \label{fig:acetac}}
\end{figure}

Considering phenomenological applications and access to different
QCD observables, hadron correlation functions are more interesting 
observables than Wilson loops. 
In order to have very good statistics and observables which do not
suffer from an intrinsically large variance, we study 
pseudo-scalar correlation functions. For cost reasons this is done
just on  $64\times32^3$ lattices.
As an example we perform a study similar to the one in \Ref{fds:quenched2}, where the
mass and decay constant of the $D_s$ as well as the charm quark mass were
investigated in the quenched approximation. Neglecting sea quark effects
allows us to generate an ensemble with the high statistics necessary for
detecting even small influences of the slow modes. However, it comes at the
price that small quark masses are not possible without running into the
problem of exceptional configurations\cite{except}. Even at the mass of the strange
quark which we take over from \Ref{fds:quenched2}, we observed at least one
clearly exceptional strange quark propagator in 40000 measurements.
We discard suspicious configurations using the criterion that the fourth moment 
of the strange pseudo-scalar correlator $M_4$
with
$
M_n = a\sum_{t=-T/2+a}^{T/2-a} t^n C_{PP}(t)
$
is at least ten standard deviations away from the average value. 

We used the 
Wilson gauge action at $\beta=6.2$ on a $64\times32^3$ lattice,
see also \Tab{t:params}, simulation C4d. 
The Wilson fermion action is non-perturbatively
$\rmO(a)$ improved 
\cite{impr:pap3} and we chose
hopping parameters\cite{fds:quenched2} $\kappa_\mathrm{strange}=0.134959$ and 
$\kappa_\mathrm{charm}=0.124637$. The quark fields are anti-periodic 
in time. 

The large statistics allows us to accurately measure 
the auto-correlation function. In \Fig{fig:acetac} we show the example of the meson 
mass with the longest auto-correlation among those considered in this study: 
the pseudo-scalar $c\bar c'$ meson with $m_{c'}=m_c=m_\mathrm{charm}$.
The normalized auto-correlation function quickly falls
to $\rho(6)\approx 0.2$, but then exhibits a long tail for which we get 
non-zero values up to $t \approx 200$. As will be discussed further in 
\sect{sec:errors}, the contribution of the slow mode to $\tauint$ is
$C_F\tauexp \approx 50\%$.

\begin{table}[!htb]
\begin{center}
\begin{tabular}[hbtp]{ccrrr}
\hline\hline
 Observable  & $\tauintl(S=1.5)$ & $\tauintl(S=3)$ & $\tauintu$ \\
\hline
 $m_\mathrm{PS}^{ss'}$  & 6.9(2) & 8.5(4)& 11(1) \\
            $f_\mathrm{PS}^{ss'}$ & 3.9(1) &4.0(1)  & 7(1) \\
            $m^{ss'}$ &  3.7(1) & 3.7(1)& 7(1) \\[2ex]
 $m_\mathrm{PS}^{cc'}$  & 11.0(4) & 13.3(7) & 15(2) \\
            $f_\mathrm{PS}^{cc'}$ & 4.3(1) & 5.0(2) & 8(1) \\
            $m^{cc'}$  & 5.6(1) & 6.6(3) & 9.4(1) \\[2ex]
 $Q^2_1$ & 183(21) & 191(31) & 196(14)\\
 $P_1$   & 12.0(4) & 13.5(7) & 15(2) \\
 $W_1(0.5\mathrm{fm},0.5\mathrm{fm})$   & 27(3) & 30(5)& 34(5) \\
\hline\hline
\end{tabular}
\end{center}
\caption{Auto-correlation times for the quenched strange and
charm quark observables along with pure gauge observables on the C4 lattice and Wilson loop on the C2 lattice. The window $W_l$ has been obtained by setting the $S$ parameter in \eq{e:UWerrW1} equal to $1.5$ and $3$. Larger values of $S$ correspond to larger windows.}
 \label{t:tauquenched}
\end{table}

Other results are shown in \tab{t:tauquenched}. The important observation
on this table is that the auto-correlation times  for all 
observables $F$ that we looked at are $\tauint(F)\lesssim 20$ 
except for the squared topological charge, for which we find
 $\tauint(Q_1^2)\approx \tauint(Q_5^2)\approx 200$. Thus there 
is very good evidence that the effect of the slow mode,
which is clearly visible in the charge,  
is strongly supressed in other observables. Still this 
supression should be verified for each new observable
and the effect of the slow mode should be estimated. The different
numbers for the $\tauint$-estimates in the table illustrate
that significant contributions by slow modes are present. 
We now turn to this issue of good error estimates.

\subsection{Improved error estimates\label{sec:errors}}

\begin{figure}[tbp!]
\begin{center}
\includegraphics{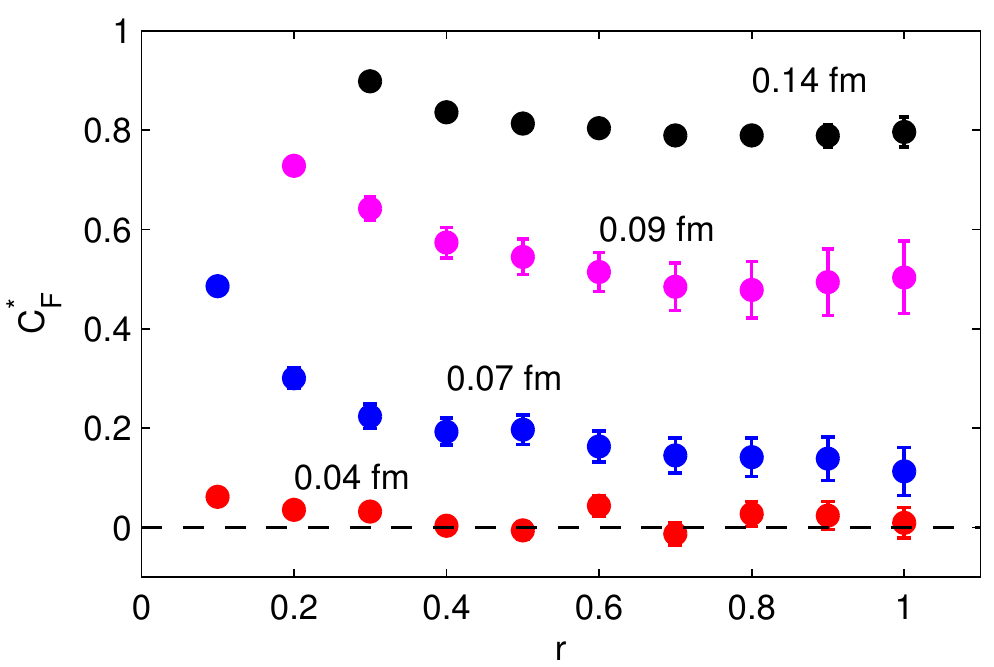}
\end{center}
\caption{The coefficient $C_{F}^{*}(r)$, \eq{e:CFstar}, for $F$ the squared, once
smeared Wilson loop of size 0.5fm$\times$0.5fm. Four different lattice spacings 
are shown in the pure gauge theory, from top to bottom: A1d, B3d, C2d, D1d. \label{f:Cstar}}
\end{figure}

Our results of the previous two subsections call for 
improved error estimates (\sect{sec:iee}) where the contribution of
a long tail of the auto-correlation functions is included. 
We discuss numerical results from both the pure gauge theory runs
and the quenched run. Also the decoupling of \sect{sec:ddcc} 
is demonstrated for these cases. 

In the pure gauge theory data we clearly see decoupling of the Wilson loops. 
Recall from \fig{f:tauintwloops} that the maximum auto-correlation is present
for the once-smeared
0.5fm$\times$0.5fm Wilson loop. In \fig{f:Cstar} we thus show  
$C_{W_1}^{*}(r)$ introduced in \Eq{e:CFstar} for that size. 
The dependence on $r$ (not to be confused with the size of the loop) is rather insignificant,
while the plot shows how the amplitude $C_{W_1}^{*}(r)$ decreases 
at smaller lattice spacings, independent of any small
residual variation with $r$. 

\begin{figure}[tbp!]
\begin{center}
\includegraphics{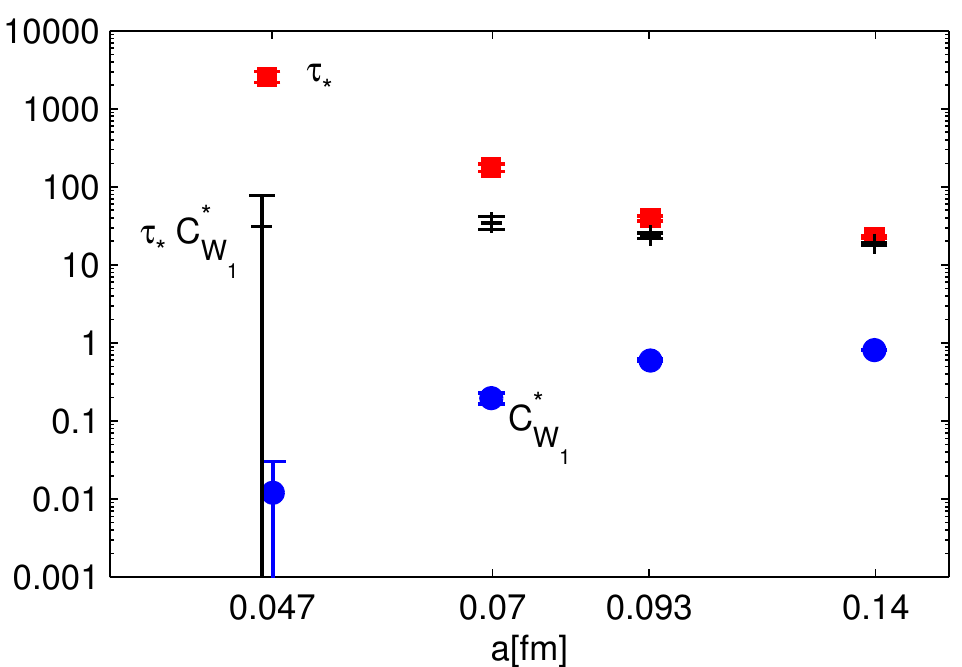}
\end{center}
\caption{The slow mode contribution $\tauexp\,C_{F}^{*}(r)$, \eq{e:CFstar}, 
for $F$ the squared, once
smeared Wilson loop of size 0.5fm$\times$0.5fm as a function 
of the lattice spacing. \label{f:Cstara}}
\end{figure}

The contribution of the slowest mode to $\tauint$ is given by the 
product $\tauexp\,C_{W_1}$.
In order to analyse the critical behavior of this 
quantity we fix $r=0.5$ and plot $\tau_*$, our estimator for $\tauexp$,
the coefficient $C_{W_1}^{*}(0.5)$ as well as the combination $\tau_*\,C^*_{W_1}$
 against the
lattice spacing in \fig{f:Cstara}. We see the strong critical slowing down
as an increase of $\tau_*$ by orders of magnitude, which is, however, 
basically compensated by the decoupling characterized by the critical behavior
of $C_{W_1}^{*}(0.5)$. As a result, the contribution of the slow mode
to the auto-correlation time of the Wilson loop stays small and 
no severe critical slowing down is observed in this quantity.

\begin{figure}[tbp!]
\begin{center}
\includegraphics{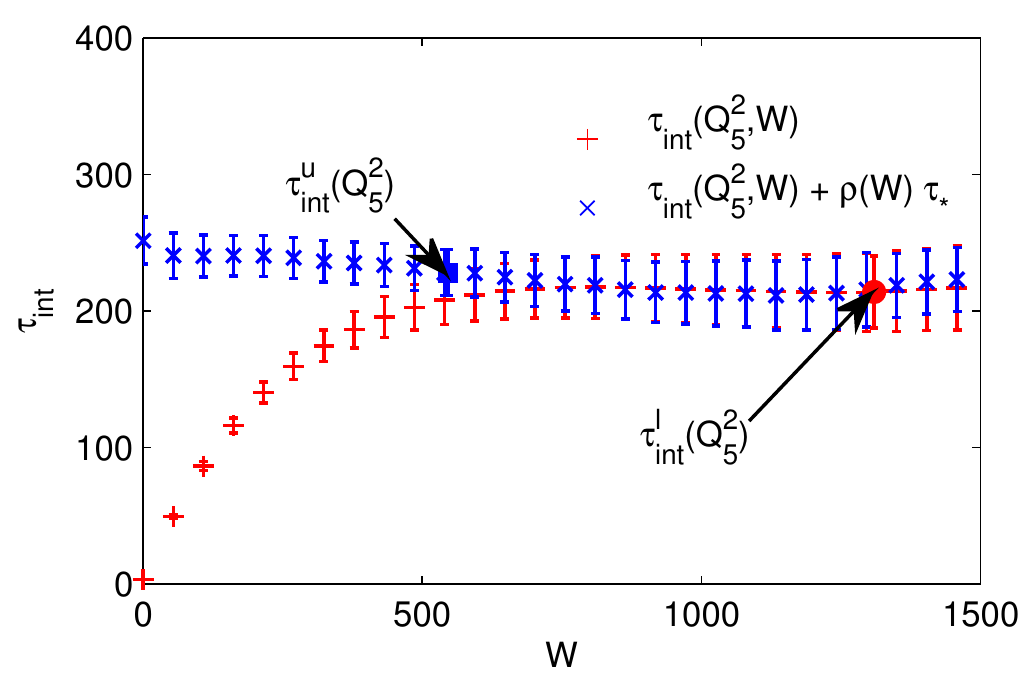}
\\
\includegraphics{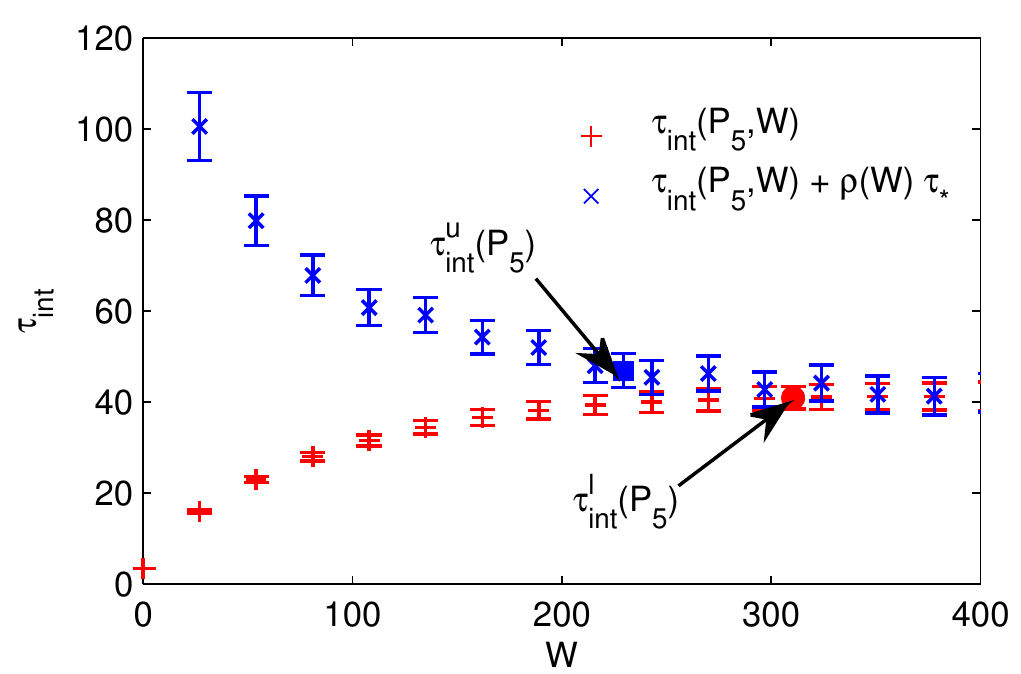}
\end{center}
\caption{Top: The error estimates $\tauintl(Q_5^2)$ and  $\tauintu(Q_5^2)$ 
  according to \Eqs{e:tauintl} and (\ref{e:tauintu}) as a function of the respective
  window $W_{l/u}$. Their values according to
  our prescription are indicated by the filled symbol points. 
  Bottom: the same for $P_5$.
  \label{f:tauintul}} 
\end{figure}

We now turn again to pseudo-scalar correlation functions on the C4d lattice. 
We saw earlier that the largest
auto-correlation is seen for the correlator of two 
distinct flavor but mass-degenerate quarks with the mass of the charm quark.
Therefore we illustrate the statements made in \sect{sec:iee} for this example as well as
for the squared topological charge. The estimate of $\tauexp$ to
be inserted into \eq{e:tauintu} and \eq{e:tauintup} was already 
discussed in \sect{sec:iee}. 
Here we show \eq{e:tauintu} in comparison to \eq{e:tauintl} 
as a function of the window size $\Wu,\Wl$. They are plotted together
in \fig{f:tauintul}. We see that $\tauintu(P_5)$ represents a much safer
estimate of $\tauint$ than $\tauintl$, also at a somewhat small
value of $\Wu$, which one might be forced to choose if the statistics is small. 
In the case of the topological charge squared, the auto-correlation
function follows closely a single exponential decay already at rather small
times. Hence the 
determination of $\tauint$ including the tail from \eq{e:tauintu} is precise 
at values of $\Wu$ which are
much smaller than the
one that is chosen by our criterion in \sect{sec:iee}.

\begin{figure}[tbp!]
\begin{center}
\includegraphics{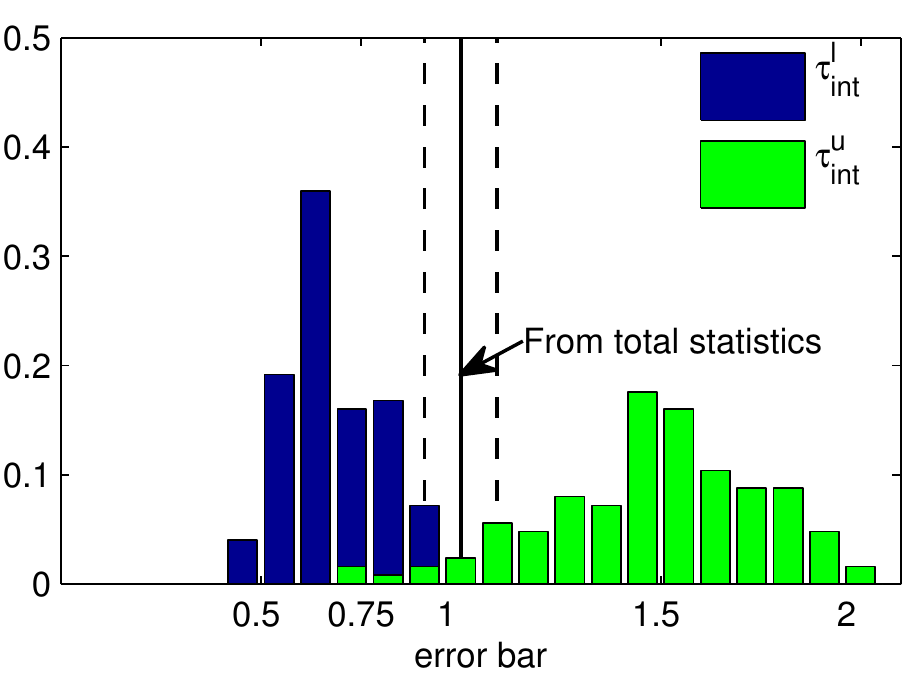}
\end{center}
\caption{Histogram distribution of the error bars (upper and lower bound) for the mass of the 
pseudo-scalar $\bar c\gamma_5 c'$. The central dashed lines show the error of the error from 
the total statistics. \label{f:Histo}}
\end{figure}

The safer error estimate described in \sect{sec:iee} is 
convincing in the case of a large statistics -- on the C4d lattice
the total statistics is around $1000\times\tauexp$. For significantly
smaller sample sizes the error estimate will of course be
less reliable. We tested the stability by dividing the total run
into pieces of about $2500\,(\mathrm{MDU}\cdot R)$ each, which is about $10\tauexp$.   
The histograms in \fig{f:Histo} show the distribution of
both standard and the improved error estimate following exactly
\sect{sec:iee}. The observable is again the $c \bar c'$ pseudo-scalar
mass. These distributions teach the following lesson.
The improved error estimate of \eq{e:tauintu} and \eq{e:tauintup}
is always safely close to the true error or somewhat above it,
while \eq{e:tauintl} with the standard window size typically underestimates
the error -- not so rarely by a factor two. An error estimate using $\tauintu$
is recommended. The histograms also remind us of an obvious fact:
typically the error of the statistical error is not that small in QCD 
simulations.

\subsection{Full QCD\label{sec:qcd}}
\begin{figure}[tbp!]
\begin{center}
\includegraphics{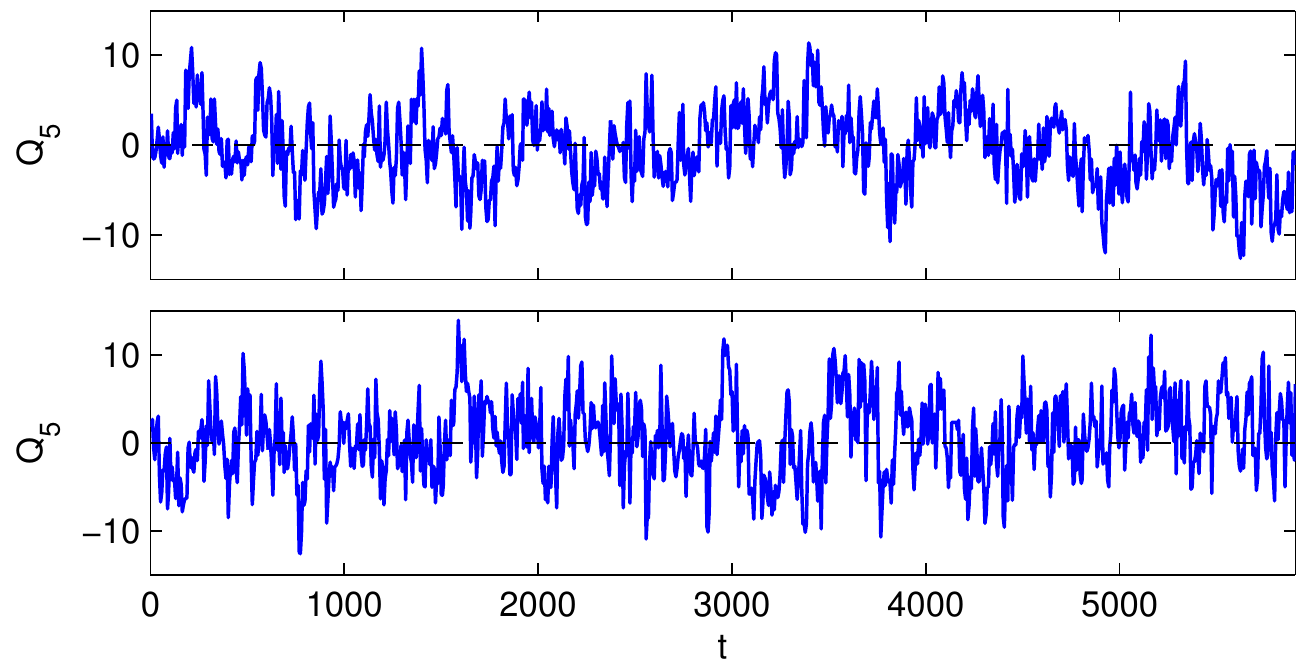}
\end{center}
\caption{Histories of the charge $Q_5$ in simulations E5f with $\tau=0.5$ (top) 
and E5g with $\tau=4$ (bottom). The Monte Carlo time $t$ is given in molecular 
dynamics time units.\label{f:historiesnf2}}
\end{figure}

As part of the CLS\footnote{\url{https://twiki.cern.ch/twiki/bin/view/CLS/}} effort,
we have carried out two rather long $\nf=2$ QCD 
runs with about 16000~MDU each, and with $R=0.37$. 
The ensemble E5f is generated with $\tau=1/2$ and 
E5g has  $\tau=4$. Both simulations describe the same physics,
using the non-perturbatively $\Oa$
improved action\cite{impr:csw_nf2} at $\beta=5.3\,,\,\kappa=0.13625$
on $64\times32^3$ lattices. The lattice spacing is read
off from  $r_0/a \approx 7 $ \cite{lat10:bjoern} and we are close to
$m_\pi r_0 = 1$, which means a pion mass of around 400~MeV. 
We will compare directly to the C1 lattice whose lattice spacing
is matched to this value of $r_0/a$. 
But first we illustrate the quality of the runs by some time histories 
in \fig{f:historiesnf2}: 
the runs contain many sign-flips of the topological charge.
As expected the frequency of topology changes is better for the 
lower, $\tau=4$, run than for the upper,  $\tau=1/2$, case.

\begin{figure}[tbp!]
\begin{center}
\includegraphics{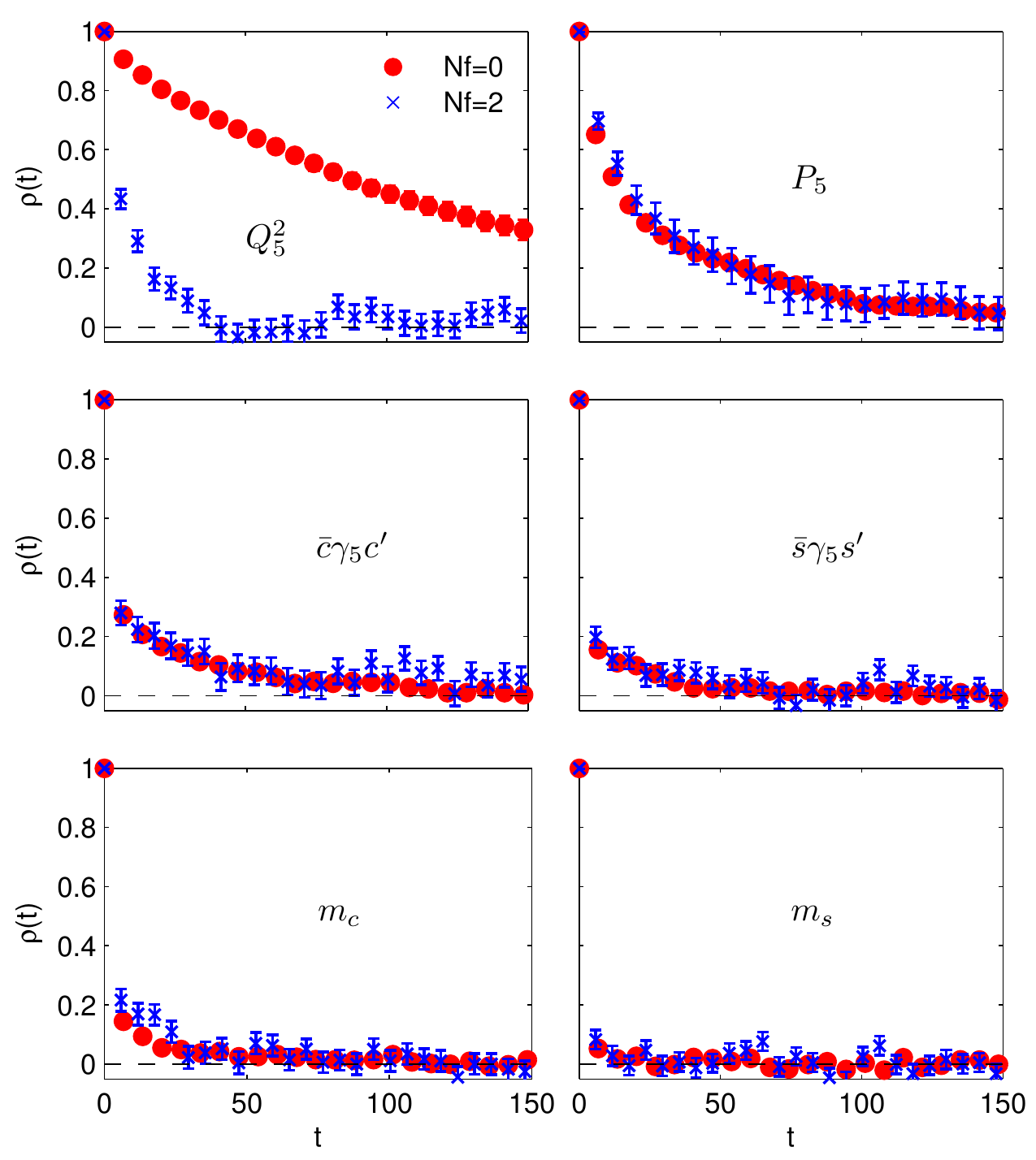}
\end{center}
\caption{Comparison of the normalized auto-correlation function $\rho(t)$ 
between quenched  and dynamical simulations at the same value of
$r_0/a\sim7$ for different observables. The data is from the C1d
and  E5g runs, respectively. 
 Top: Comparison for topological charge squared and plaquette. Center: 
 Pseudo-scalar meson masses with mass-degenerate quarks of the charm quark 
 mass on the left ($\bar c\gamma_5 c'$) and  strange quark mass on 
 the right ($\bar s\gamma_5 s'$), extracted from 
plateau averages over $x_0\in[23a,27a]$. Bottom: Auto-correlation functions of
PCAC quark masses at $x_0 = 24a$.  \label{f:rhoE}}
\end{figure}

A one-to-one comparison of the auto-correlation functions, quenched vs. 
$\nf=2$,  is presented 
in \fig{f:rhoE}. One observes a very similar decorrelation of all observables
quenched and in full QCD, except for the squared topological charge
which decorrelates much faster with dynamical fermions. Unfortunately
we cannot offer a real theoretical understanding of this rather striking
observation. However,
note that the change of the gauge action (in the pure gauge theory) 
from Wilson plaquette action to Iwasaki action has a similar effect,
namely the auto-correlations of $Q^2$ were strongly affected while 
auto-correlations of other observables are essentially unchanged. 
Among the effects of the introduction of the fermion determinant is 
a change of the effective gauge action in the ultraviolet. Beyond the
leading $\beta$-shift there are also dimension six terms and this 
``part'' of the fermion determinant is
the same as a change of the lattice gauge action.

\begin{figure}[hbtp!]
\begin{center}
\includegraphics{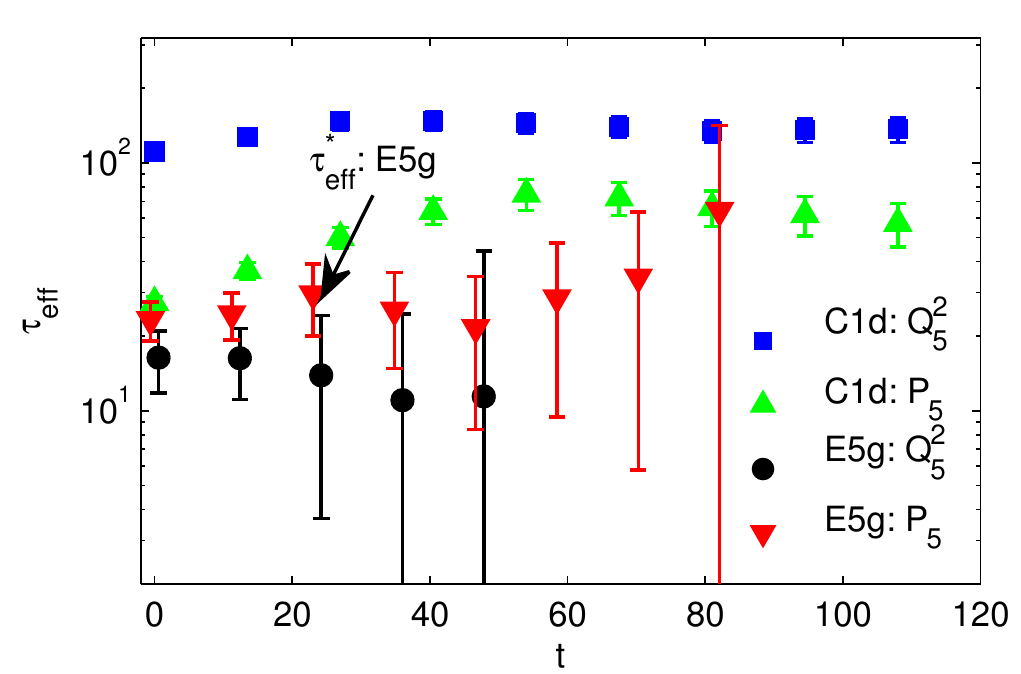}
\end{center}
\caption{Estimators for the exponential auto-correlation time
from smeared plaquette and topological charge in pure gauge (C1d) and 
dynamical (E5g) simulations.\label{f:tauexpE}}
\end{figure}

As we did for the pure gauge theory, we now come to
the extraction of the exponential
auto-correlation time, see \fig{f:tauexpE}. The estimator 
$\tauexp^{\mathrm{eff}}(t,F) =  \frac{t}{2 \log\left\{{\rho_{F}(t/2) \over \rho_{F}(t)}\right\}}$
is shown for those observables with the slowest decorrelation out
of our set. Clearly the determination of $\tauexp^{\mathrm{eff}}$ is difficult, 
but it seems possible. The location of  $\tau_*$, which we remind the reader
is our estimate \eq{e:taustar} for $\tauexp$, is indicated in the figure.

The numbers
for $\tauint$ and $\tau_*$ are listed in \tab{t:tauintnf2}. 
We see again that auto-correlation times
for long trajectories with length $\tau=4$ are around a factor two
smaller than those for $\tau=1/2$. 
In the table we list numbers for $\tau_*$ determined {\em just} from the 
indicated observable for illustration. In our estimate of $\tauintu$ the 
maximum one is then taken into account as defined in \eq{e:tauexpeff}.
The more observables one considers, the better (larger) the estimate of $\tauexp$ will
get. Even if this is still below the true value of $\tauexp$,
it will provide us with a more realistic estimate of $\tauint$.

\subsubsection{Proposal for error estimates}

The numbers in \tab{t:tauintnf2} come from a rather long simulation. 
Such data is not always available. Here we propose how one may proceed
in this situation, using a reasonable estimate of the contribution of the 
tail of the auto-correlation function.  
\Eq{e:tauintu} should be used when an onset of a tail is 
visible in the data and we suggest to choose $\Wu$ such that 
$\delta[\rho_F(\Wu)] \approx \rho_F(\Wu)/3$.
On the other hand for auto-correlations which quickly fall off,
\eq{e:tauintup} is recommended. With low statistics 
an estimate of $\tauexp$, needed for these formulae, is
impossible to obtain. We therefore suggest to use
a value for a not-so-small $a$ with good statistics
together with the scaling observed in the 
pure gauge theory. For our 
$\Oa$-improved action at $\nf=2$ we hence suggest
an $a^{-5} \sim \exp(7\beta)$ scaling 
(see eq.(4.5) of \cite{lat07:rainer}). 
Together with 
$\tauexp\approx 40$ at $\beta=5.3$, this leads 
for the action of \cite{impr:csw_nf2} to\footnote{To
exclude any confusion, we here put the units explicitly 
which we have been using throughout.}
\be \label{e:tauexpmodel}
  R\,\tauexp\approx 200\,\exp(7\,(\beta-5.5))\,\mathrm{MDU} \ .
\ee
For safety reasons, one may attach an error of a factor 2 
to this estimate and should of course be aware of the assumptions
made above. The best situation is an observable with 
a strong decoupling, i.e.
a small auto-correlation function $\rho_F(\Wu)$ or $\rho_F(W_0)$, 
for which the intrinsic uncertainties of the
model \eq{e:tauexpmodel} are not that relevant.

\begin{table}
\begin{center}
{\small
\begin{tabular}[hbtp]{l | lllll | lll}
\hline\hline
 & \multicolumn{5}{|c|}{$Q_5^2$} &  \multicolumn{3}{|c}{$P_5$}\\
 \hline
TAG & $\langle Q_5^2 \rangle $ & $a^4\chi_\mathrm{t} $ & $\tauintl$ & $\tauintu$ & $\tau_* $ &  $\tauintl$ & $\tauintu$ & $\tau_* $ \\ 
 \hline
 C1d & 50(4)  & 2.4(2) $ \times 10^{-5}$  & 137(25) & 134(15) & 140(18) & 38(4)  & 43(5) & 74(11)  \\
 E5f & 17.3(1.8)  & 0.82(9) $ \times 10^{-5}$  & 16(4)   & 23(5)   & 21(5)   & 84(31) & 66(13) & 66(19) \\
 E5g & 18.9(1.5)  & 0.90(7) $ \times 10^{-5}$  & 10(2)   & 14(3)   & 15(4)   & 29(8)  & 29(5) & 39(12) \\
 \hline\hline
 \end{tabular}
}
\end{center}

\caption{Comparison between the integrated auto-correlation times for the 
topological charge and the smeared plaquette between the dynamical and the corresponding
pure gauge ensemble.  The dynamical runs E5f and E5g have trajectory length 
$\tau=0.5$ and $\tau=4$, respectively. The pure gauge run C1d has $\tau=4$. 
Also the value of the topological susceptibility, $\chi_\mathrm{t}$, is given.\label{t:tauintnf2}
}
\end{table}

\section{Summary and conclusions}
In this paper we have established a very severe critical slowing down
of the topological charge in pure Yang-Mills theory when using the (DD)-HMC
algorithm. A dynamical critical exponent of $z=5$ means that, 
at constant volume,
the full simulation 
scales with $a^{-10}$ at least, since an HMC type algorithm is 
expected to scale with $a^{-4}$ from the increasing number
of lattice points and typically an additional factor of $a^{-1}$ 
from the decreasing step size. However, we also investigated 
Wilson loops, which are more commonly
in the focus of interest. They are not affected in the same way, 
exhibiting a much milder
slowing down while approaching the continuum limit ($z\approx 0.5\ldots 1$).
Martin L\"uscher investigated observables after integrating
the Wilson flow\cite{Wflow} for some distance which
removes UV fluctuations. These observables
effectively show $z\approx2$ \cite{lat10:martin},
a critical slowing down in between $Q^2$ and the Wilson loops.

We have also considered observables formed from
pseudoscalar correlation functions, both in a quenched setting
and for $\nf=2$. These quantities are of immediate interest 
and at the same time not plagued by large UV fluctuations. At a lattice
spacing of $a\approx0.07$fm their autocorrelation functions are much better 
behaved than the one of $Q_5^2$ but a weak coupling to the slow mode is seen
in \fig{f:rhoE}. 
Unfortunately, a systematic study as a function of the 
lattice spacing and quark mass is prohibitively expensive,
but we expect that these observables continue to couple only weakly
to the slow mode 
and their slowing down is significantly less severe 
than for the topological charge.

On the one hand, this is encouraging. In practical simulations, unless
we are interested in the slow observables themselves,
we do not need to gather enough statistics to accurately determine
their auto-correlation time. It is sufficient to have a decent sampling in
the slow modes to assure practical ergodicity, i.e. a few times their
auto-correlation times is needed. On the other hand, 
the danger remains that
there are even slower modes which are so slow that the corresponding
fluctuations do not show in the full runtime. The only way to study
this is to start simulations in parameter space, which can be considered
safe and then move in small steps towards the critical points, 
monitoring a large number of observables and relying on the continuity
of auto-correlations in terms of the system's parameters. 

Even if the coupling to the slow modes may be small 
it is important for a correct error analysis. We described
a practical method to take these effects into account. 
It relies on the fact that information about $\tauexp$ can be obtained
through observables which couple strongly to the corresponding
mode. Under these circumstances, the error analysis can be made
significantly safer.

Still, a true solution to the critical slowing down has to be an 
algorithmic one
which at least solves the problem regarding the topological charge. The
dramatic progress in the fermion algorithms which the field has witnessed
 during the last 
decade gives us hope that this can actually be achieved.

\noindent
{\bf Acknowledgements.}
We would like to thank M. L\"uscher and F. Palombi
for many useful discussions and the possibility to use some of their data
as well as U. Wolff and other members of CLS for helpful  discussions.
We are grateful to A. Kennedy for a critical reading of an 
earlier version of the manuscript.
This work is supported by the Deutsche Forschungsgemeinschaft
in the SFB/TR~09
and by the European community
through EU Contract No.~MRTN-CT-2006-035482, ``FLAVIAnet''.
We thank the John von Neumann institute for computing and
the HLRN 
for allocating computer time for this project. Part of our runs 
were performed on the PAX cluster at DESY, Zeuthen.

\begin{appendix}

\section{Error of the Error }
\label{sec:ee}
An introduction to the 
error analysis of correlated data from a Markov chain 
with references
can be found in \cite{Sokal} while the case of functions of the primary 
observables is treated in \cite{UWerr}. 
Here we review the main formulae for
estimating the error of the error from these references
and make those explicit which are not given there.

Our estimator for the mean, \eq{e:mean}, and the auto-correlation function,
\eq{e:gammadef}, are
\bea
 \Obar_\alpha &=&{1\over N} \sum_{t=1}^N O_\alpha(q_t)\;,\quad \Obar_F = F(\Obar_\alpha)\\
 \Gammabar_{FF'}(t) &=&  
 {1\over N-|t|}\sum_{u=1}^{N-|t|} dF(q_u) d{F'}(q_{u+t}) + 
 {1\over N^2}\sum_{u=1}^{N} dF(q_u) d{F'}(q_{u}) \,,\\
\text{where}&&\nonumber\\
dF(q) &=& \sum_\alpha F_\alpha (O_\alpha(q) - \Obar_\alpha)
\eea
which for $\Gammabar_{FF'}$ contains a bias correction discussed in \cite{UWerr}.
In the computation of our auto-correlation times, e.g. \eq{e:tauintu}, 
we replace $\Gamma$ by its estimator $\Gammabar$.

In \sect{sec:2} we introduced various quantities which
are functions $G(\{\Gamma_F(t)\})$. Their error 
is computed from simple error propagation
\be\label{e:errorG}
 (\delta G)^2 = {1\over N} \sum_{A,B}{\partial G\over\partial\Gamma_A}\Sigma_{AB}{\partial G\over\partial\Gamma_B}
\ee
where $A=(F,u)$ and $B=(F',v)$ collect the observable
and time variable and run over all components of the
auto-correlation functions. The covariance matrix $\Sigma_{AB}$
is given by
\bea
\label{e:cov4moment}
\Sigma_{AB} 
&=&{1\over N^2}\sum_{s=1}^{N}\sum_{t=1}^{N}
 dF(q_s) dF(q_{s+u}) d{F'}(q_t) d{F'}(q_{t+v})
 - {\Gammabar_F}(u){\Gammabar_{F'}}(v)\;.
\eea 
Neglecting the completely connected part of the fourth moment in 
\eq{e:cov4moment}, we have the approximation
\be\label{e:covMatrix}
\Sigmabar_{AB} \approx {1\over N}\sum_{m=-\infty}^{\infty}\Gamma_{FF'}(m+v)\Gamma_{FF'}(m+u)+\Gamma_{FF'}(m+v)\Gamma_{FF'}(m-u)\,,
\ee
which in practice we evaluate with the cut \eq{e:GammaTilde} 
to be discussed below.

In some simple cases, approximations can be  applied to \eq{e:errorG}
to derive more compact error 
formulae, for example \cite{Madras,algo:L2}
\begin{align}
(\delta{\tauint(W)})^2&\approx {2(2W+1)\over N }\tauint^2(W)\;,\\
(\delta{\rho(t)})^2&\approx {1\over N}\sum_{m=1}^\infty(\rho(m+t)+\rho(m-t)-2\rho(m)\rho(t))^2\,.\label{e:errorML}
\end{align}
Furthermore we used the approximation
\be
(\delta {\tauintu})^2\approx (\delta{\tauint(W_u)})^2 
 + \tauexp^2 (\delta{\rho(W_u)})^2 + \rho^2(W_u)(\delta{\tauexp})^2\;,
\ee
for the error of \eq{e:tauintu} 
after checking that the left out cross terms are negligible. 
Also in the case of \eq{e:CF2} a similar approximation has been applied.
For all other quantities we remained directly with \eq{e:errorG}.

As mentioned, the evaluation of the sum in \eq{e:covMatrix} and also the one in \eq{e:errorML}
require the introduction of a cutoff on the sum over $m$.
This can be done with the function
\be\label{e:GammaTilde}
\tilde{\Gamma}(x)=
  \begin{cases}
    \Gammabar(x) & |x| \leq W_\Sigma \\
    0         & |x| > W_\Sigma \;.
  \end{cases}
\ee
(and $\tilde{\rho}$) where we omit the subscripts to keep notation light.
The computation of $\Sigma(u,v)$  is then carried out as the sum of 
terms 
$\sum_m \tilde{\Gamma}(m)\tilde{\Gamma}(m+t)$ at times $t=u+v$ and $t=u-v$:
\be
\Sigmabar(u,v)\approx{1\over N}\sum_m\left( \tilde{\Gamma}(m)\tilde{\Gamma}(m+u+v)+\tilde{\Gamma}(m)\tilde{\Gamma}(m+u-v)\right)
\ee
that can be done in at most $O(W_\Sigma^2)$ operations.
The size of $W_\Sigma$ can be discussed in similar manner as we have done about 
window sizes for the auto-correlation function. 
In our analysis we take $W_\Sigma = W$, with $W$ given by prescription \eq{e:UWerrW}, 
however, using  $W_\Sigma = 2W$ gives similar results.

In principle one could think about adding a tail contribution to 
\eq{e:GammaTilde}, but this goes too far, even with the statistics 
we had at hand.

\end{appendix}

\bibliographystyle{JHEP}   
\bibliography{topo,latticen}           


\end{document}